\documentclass[preprint,nofootinbib,aps,superscriptaddress,eqsecnum]{revtex4-1}
\usepackage{graphicx}
\usepackage{mathrsfs}
\usepackage{soul}
\usepackage{natbib}
\usepackage{enumerate}
\usepackage{amsmath,amssymb}
\usepackage{slashed}
\usepackage{amssymb,amsmath,subfigure,xcolor}

\newcommand{\lapprox}{%
\mathrel{%
\setbox0=\hbox{$<$}
\raise0.6ex\copy0\kern-\wd0
\lower0.65ex\hbox{$\sim$}
}}
\newcommand{\gapprox}{%
\mathrel{%
\setbox0=\hbox{$>$}
\raise0.6ex\copy0\kern-\wd0
\lower0.65ex\hbox{$\sim$}
}}

\newcommand{\ba}{\begin{array}}
\newcommand{\ea}{\end{array}}
\newcommand{\bd}{\begin{displaymath}}
\newcommand{\ed}{\end{displaymath}}
\newcommand{\beq}{\begin{equation}}
\newcommand{\eeq}{\end{equation}}
\newcommand{\bea}{\begin{eqnarray}}
\newcommand{\eea}{\end{eqnarray}}
\newcommand{\Z}{\mathbb{Z}}

\newcommand{\nn}{\nonumber}
\newcommand*\DAlambert{\mathop{}\!\mathbin\Box}
\def\etc{ {\em etc.\ }}
\def\ie{ {\em i.e.,\ }}

\def\a{\alpha}

\def\b{\beta}
\def\g{\gamma}

\def\l{\lambda}
\def\m{\mu}
\def\n{\nu}

\def\q2 {q^2}

\def\bt{\begin{table}}
\def\et{\end{table}}
\catcode`@=11 % This allows us to modify PLAIN macros.
\def \gsim{\mathrel{\mathpalette\@versim>}}
\def \lsim{\mathrel{\mathpalette\@versim<}}
\def \@versim#1#2{\lower0.4ex\vbox{\baselineskip\z@skip\lineskip\z@skip
     \lineskiplimit\z@\ialign{$\m@th#1\hfil##\hfil$%
     \crcr#2\crcr\sim\crcr}}}
\catcode`@=12 % at signs are no longer letters

\begin{document}

%\begin{flushright}
%{\small CUPP }
%\end{flushright}
\renewcommand*{\thefootnote}{\fnsymbol{footnote}}

\begin{center}

{\large\bf Higher dimensional operators in 2HDM}\\[15mm] 
Siddhartha Karmakar\footnote{E-mail: phd1401251010@iiti.ac.in} and Subhendu
Rakshit\footnote{E-mail: rakshit@iiti.ac.in} 
\\[2mm]

{\em Discipline of Physics, Indian Institute of Technology Indore,\\
 Khandwa Road, Simrol, Indore - 453\,552, India}
\\[20mm]
\end{center}

\begin{abstract} 
\vskip 20pt
We present a complete (non-redundant) basis of CP- and flavour-conserving six-dimensional operators in a two Higgs doublet model (2HDM).  We include  $\Z_{2}$-violating operators as well. In such a 2HDM effective field theory (2HDMEFT), we estimate how constraining the 2HDM parameter space from experiments can get disturbed due to these operators. Our basis is motivated by the strongly interacting light Higgs (SILH) basis used in the standard model effective field theory (SMEFT). We find out bounds on combinations of Wilson coefficients of such operators from precision observables, signal strengths of Higgs decaying into vector bosons\etc In 2HDMEFT, the 2HDM parameter space can play a significant role while deriving such constraints, by leading to reduced or even enhanced effects compared to SMEFT in certain processes. We also comment on the implications of the SILH suppressions in such considerations. 
\end{abstract}

\vskip 1 true cm
\maketitle

%\newpage
\setcounter{footnote}{0}
\renewcommand*{\thefootnote}{\arabic{footnote}}

%\def\baselinestretch{1.5}
%==============================================================================
%==============================================================================
\section{Introduction}
Two Higgs doublet model is the most studied extension of the scalar sector of the Standard model (SM) of particle physics. Inclusion of an additional scalar doublet is necessary in the supersymmetric models. There are other phenomenological  motivations for considering non-supersymmetric versions of this model as well. For example, an additional Higgs doublet can lead to  a successful electroweak baryogenesis~\cite{Trodden:1998ym}. Moreover,  anomalies in tauonic B-decays can be addressed in a 2HDM~\cite{Crivellin:2012ye, Celis:2012dk,Crivellin:2015hha}.  

 Non-observation of new fundamental particles at the LHC motivates us to formulate the SM effective Lagrangian below a TeV or so. The recent data implies that the newly-discovered $125$ GeV  Higgs boson is SM-like, but still allows for the existence of new scalars at sub-TeV scales. In the alignment limit, 2HDM can accommodate such new scalars so that the extra contributions coming from the renormalisable 2HDM Lagrangian to various processes involving SM particles are reasonably small.  This calls for an extensive study of higher dimensional operators in exploring 2HDM phenomenology as their effects can be of the same order of the extra contribution in 2HDM at the tree-level in the alignment limit.

In this paper, we formulate a basis of independent six-dimensional~(6-dim) operators assuming 2HDM to be the low-energy theory.  We include the $\Z_2$-violating operators, but, for simplicity, exclude  the CP- and flavour-violating ones. The bounds imparted on the Wilson coefficients by the electroweak precision tests (EWPT) are estimated. Such operators also affect the signal strengths of the SM-like Higgs boson decaying into a pair of vector bosons. We estimate such contributions as in future, better measurements of these signal strengths can further tighten constraints on these Wilson coefficients. Our results also reflect the fact that the inclusion of higher dimensional operators can relax the bounds on the parameter space of renormalisable 2HDM. We also discuss the cases where we have weighed the Wilson coefficients following the SILH prescription as in the SMEFT~\cite{Giudice:2007fh} and mark the resulting changes in the above-mentioned bounds.

The study of a complete set of higher dimensional operators of SM goes long back~\cite{Buchmuller:1985jz,Leung:1984ni}, where the authors formulated the basis of 6-dim operators of SM assuming lepton and baryon number conservation. A systematic study of the electroweak precision constraints on the Wilson coefficients of the bosonic 6-dim operators was first done in the framework of the so-called HISZ basis of SMEFT~\cite{Hagiwara:1993ck}. The problem of writing a complete set of 6-dim operators of SMEFT was revisited in Ref.~\cite{Grzadkowski:2010es} where the equations of motions (EoM) of all the fields were used to identify 21 redundant operators in the basis of Ref.~\cite{Buchmuller:1985jz}. The basis for SMEFT introduced by the latter is often referred to as the Warsaw basis. The SILH basis~\cite{Giudice:2007fh} was formulated in the context of scenarios where the hierarchy problem is alleviated due to the existence of a strongly-interacting sector beyond the TeV scale. There are two broad classes of models which can be represented by the SILH Lagrangian. First, the extra-dimensional models where the Higgs boson is a part of the bulk and the rest of the SM fields are part of a brane at low-energy~\cite{Agashe:2004rs}. The other one consists of all the models where the Higgs is a pseudo Nambu-Goldstone boson\,(pNGB) of the strong sector. The Little Higgs~\cite{ArkaniHamed:2002qy} is an archetype of the latter class. The composite Higgs models can be a part of both the classes. The SILH Lagrangian is described by two scales of new physics, namely $f$, the compositeness scale and $m_{\rho}\sim g_{\rho} f$, the lightest vector boson mass in the strongly-interacting sector, with $g_{\rho}$ being the coupling of new strong sector. An updated review of the study of SMEFT in light of precision electroweak and Higgs signal strength data can be found in Refs.~\cite{Falkowski:2015fla,Brivio:2017vri}.

The study of a composite 2HDM  in the context of a $SO(6)/SO(4)\times SO(2)$ coset was performed in Ref.~\cite{Mrazek:2011iu}.  Perturbative unitarity bounds on S-matrix elements were extracted and collider phenomenology of this particular model has been explored only recently~\cite{DeCurtis:2016scv,DeCurtis:2016gly,DeCurtis:2017gzi}. The 2HDM in a Little Higgs scenario was studied in Refs.~\cite{Brown:2010ke,Gopalakrishna:2015dkt}. A few other studies of the composite inert doublet model of dark matter have been carried out as well~\cite{Fonseca:2015gva,Carmona:2015haa}. 

In Ref.~\cite{Chala:2017sjk}, an extension of SMEFT SILH basis incorporating a light singlet scalar along with SM degrees of freedom was introduced. Impact of some of the six-dimensional operators involving two Higgs doublets on exotic decay channels of charged Higgs boson were studied in Ref.~\cite{DiazCruz:2001tn}. The kinetic terms comprising of  four scalars and two derivatives  in a N-Higgs doublet scenario were studied recently in Refs.~\cite{Kikuta:2015pya,Kikuta:2011ew,Kikuta:2012tf}.  An attempt to write down the full set of 6-dim operators in 2HDM was made in Ref.~\cite{Crivellin:2016ihg} in a Warsaw-like basis. In contrary, our basis is motivated by the SILH basis in SMEFT and is a complete one, as  we point out that there is a redundant operator in the basis of Ref.~\cite{Crivellin:2016ihg}. In addition, as we mentioned earlier, we include the $\Z_2$-violating operators as well.

We start with introducing the tree level Lagrangian in 2HDM and the corresponding equations of motion and then formulate our basis in Section~\ref{sec2}. In Section~\ref{scalars}, we carry out the kinetic and mass diagonalisations of the scalar fields which enable us to write down the effective couplings of those scalars with the vector bosons. Section~\ref{EWPT} deals with the constraints on the Wilson coefficients coming from the EWPT. In section~\ref{section:pheno}, we evaluate the decay widths and signal strengths of the SM Higgs boson decaying into vector boson pairs. 
%Section VI is dedicated to the implication of EFT operators on the alignment limit of 2HDM. 
Finally in Section~\ref{sec:conclude} we consolidate our results  and eventually conclude.   

\section{Construction of the 2HDMEFT}\label{sec2}

\subsection{The 2HDM Lagrangian and classical EoMs}
\label{2hdm}
We use the same notation as in Ref.~\cite{Crivellin:2016ihg} in order to avoid further confusion. The Higgs fields in the doublet notation  can be written as: 
\begin{equation} 
\varphi_{I} =  \left(
\begin{array}{c}
\phi_{I}^{+}\\
\frac{1}{\sqrt{2}}(v_{I} + \rho_{I}) + i\, \eta_{I}\\
\end{array}
\right)
\end{equation}
where $I = 1,2$. Before the two Higgs fields $\varphi_{1}$, $\varphi_{2}$ get vacuum expectation values (vev) the renormalisable 2HDM Lagrangian is given by: 
\begin{multline}
\mathcal{L}_{2HDM}^{(4)} = -\frac{1}{4} G_{\mu\nu}^a G^{a \mu\nu} - \frac{1}{4} W^{i}_{\mu\nu}W^{i\mu\nu} - \frac{1}{4} B_{\mu\nu}B^{\mu\nu} + |D_{\mu}\varphi_{1}|^2 + |D_{\mu}\varphi_{2}|^2 - V(\varphi_{1},\varphi_{2})  \\ + \textit{i} \, (\bar{q}\slashed{D} q + \bar{l}\slashed{D} l + \bar{u}\slashed{D} u + \bar{d}\slashed{D} d) + \mathcal{L}_{Y}, 
\end{multline}
where the first three are field strengths of the gauge bosons of $SU(3)_{C}$, $SU(2)_L$ and $U(1)_Y$ respectively. The indices $a = 1,..,8$ and $i = 1,2,3$ are summed over. The tree-level 2HDM potential is given by: 
\begin{multline}
\label{potential}
V(\varphi_{1},\varphi_{2}) = m_{11}^2 |\varphi_{1}|^2
 + m_{22}^2 |\varphi_{2}|^2 - ( m_{12}^2 \varphi_{1}^{\dagger} \varphi_{2} +  h.c.) + \lambda_1 |\varphi_{1}|^4 + \lambda_2 |\varphi_{2}|^4 + \lambda_{3} |\varphi_{1}|^2 |\varphi_{2}|^2\\
  +  \lambda_4 |\varphi_{1}^{\dagger} \varphi_{2}|^2 + \frac{\l_5}{2} ( (\varphi_{1}^{\dagger} \varphi_{2})^2 + h.c.)  + (\l_6 |\varphi_{1}|^2 + \l_7 |\varphi_{2}|^2)( \varphi_{1}^{\dagger} \varphi_{2} + h.c.).
\end{multline}
The term with coefficient $m_{12}^2$ is called the soft $\Z_{2}$-violating term, whereas the ones with $\l_6$ and $\l_7$ are called the hard $\Z_2$-violating terms as they give rise to quadratically divergent contribution to $\varphi_{1}-\varphi_{2}$ mixing. However, one is allowed to start with non-zero values of $\l_6$ and $\l_7$   as long as they can be rotated to $\l_6, \l_7 = 0$ using reparametrisation transformations~\cite{Ginzburg:2004vp,Ginzburg:2008kr}. This scenario is referred to as ``hidden soft $\Z_2$-violation". Moreover, in SILH scenarios, one considers the existence of a strongly-interacting sector at $\sim \mathcal{O}$(1~TeV) which deliver Higgs as a pNGB at low energy. In those cases new resonances at or above $\sim \mathcal{O}(1$~TeV) take care of the quadratic divergence of Higgs mass, solving the hierarchy problem. The same mechanism will take care of the quadratic divergence in $\varphi_1 - \varphi_{2}$ mixing caused by $\l_6$ and $\l_7$ for the 2HDMs which are governed by such a strongly-coupled sector at higher energies. That is why in this paper we carry out all calculations keeping $\l_6, \l_7 \neq 0$. The same explanation holds true for  the inclusion of $\Z_{2}$-odd higher dimensional operators.  

The general Yukawa Lagrangian is given by,
\begin{equation}
\mathcal{L}_{Y} = -\sum_{I=1,2} Y^{e}_I \, \bar{l} \,e \varphi_{I} - \sum_{I=1,2} Y^{d}_I \, \bar{q} \, d \varphi_{I} - \sum_{I=1,2} Y^{u}_I \, \bar{q} \, u \tilde{\varphi}_{I}.
\end{equation}

For eliminating the redundant operators from the basis of 6-dim operators, one needs to derive the EoMs of the bosonic fields from the tree-level 2HDM Lagrangian. It is necessary to separate out the redundant ones because they do not contribute to the S-matrix elements~\cite{Politzer:1980me}. While doing that, we neglect the five-dimensional operators~\cite{Grzadkowski:2010es}. The EoMs are given as:
\begin{eqnarray}
\DAlambert \varphi_{1}^{i} &=& - m_{11}^2 \varphi_1^{i} - m_{12}^2 \varphi_2^{i} - 2 \l_1 |\varphi_{1}|^2 \varphi_{1}^{i} - \l_3 |\varphi_{2}|^2 \varphi_{1}^{i} - \l_4 (\varphi_{2}^{\dagger} \varphi_{1}) \varphi_2^{i} 
- \l_5 (\varphi_1^{\dagger} \varphi_{2}) \varphi_2^{i} \nn \\  
&&-(( \l_6 \varphi_1^{\dagger} \varphi_{2} +  \l_6^{*} \varphi_2^{\dagger} \varphi_{1}) \varphi_1^{i} +  \l_6 |\varphi_1|^2 \varphi_2^{i}) - \l_7 |\varphi_2|^2 \varphi_2^{i} - Y_{1}^{d \dagger} \bar{d} q^{i} - Y_{1}^{e \dagger} \bar{e} l^{i} + Y_1^{u} \epsilon^{ij} \bar{q}^{j} u, \nn\\
\DAlambert  \varphi_{2}^{i} &=& - m_{22}^2 \varphi_2^{i} - {m_{12}^{2*}} \varphi_2^{i} - 2 \l_2 |\varphi_{2}|^2 \varphi_{2}^{i} - \l_3 |\varphi_{1}|^2 \varphi_{2}^{i} - \l_4 (\varphi_{1}^{\dagger} \varphi_{2}) \varphi_1^{i} - \l_5 (\varphi_2^{\dagger} \varphi_{1}) \varphi_1^{i} \nn\\ 
&&-\l_6 |\varphi_1|^2 \varphi_1^{i} - ((\l_7 \varphi_1^{\dagger} \varphi_{2} + \l_7^{*} \varphi_2^{\dagger} \varphi_{1}) \varphi_2^{i} + \l_7 |\varphi_2|^2 \varphi_1^{i}) - Y_{2}^{d \dagger} \bar{d} q^{i} - Y_{2}^{e \dagger} \bar{e} l^{i} + Y_2^{u} \epsilon^{ij} \bar{q}^{j} u, \nn\\
%\end{multline}
%\begin{equation}
\partial^{\rho} B_{\rho \mu} &=& g' \Big( \sum_{I = 1,2} Y_{\varphi{I}} \, \varphi_{I}^{\dagger} \, \textit{i} \overset\leftrightarrow{D_{\m}} \, \varphi_I +  \sum_{\psi = q,l,u,d,e} Y_{\psi} \bar{\psi} \, \gamma_{\m} \psi \Big), \nn\\
D^{\rho} W^{i}_{\rho \mu} &=& \frac{g}{2} \Big(    \sum_{I = 1,2}  \varphi_{I}^{\dagger} \, \textit{i} \overset\leftrightarrow{D_{\m}} \, \varphi_I + \bar{l} \, \g_{\m} \, \tau^{i} \, l + \bar{q} \, \g_{\m} \, \tau^{i} \, q \Big). 
\end{eqnarray}

\subsection{Operator basis}
\label{operators}

In the universal theories~\cite{Barbieri:2004qk}, the deviations of the properties of the Higgs boson from SM can be expressed in terms of only the higher-dimensional bosonic operators. Both the Warsaw basis~\cite{Grzadkowski:2010es} and SILH basis~\cite{Giudice:2007fh} are bosonic bases,\ie all bosonic operators are kept in those bases. The effects of the 14 bosonic operators on Higgs physics were discussed in context of the SILH basis~\cite{Elias-Miro:2013mua}, where the RG evolutions of their Wilson coefficients were also studied. It was pointed out that the 14 operators capture all the new physics effects of the Higgs sector in the SILH basis; but it takes more than 14 operators to express the same effects in the Warsaw basis. Moreover, the study of the RG analysis of the Wilson coefficients also implied that in SILH basis, the tree-level and loop-level operators do not mix under running, which is not the case for the Warsaw basis. In principle, all the bases are equivalent if they are complete and non-redundant. However, the new physics effects in the Higgs sector are expressed with a fewer number of operators in the SILH basis compared to the Warsaw one. This gives the SILH basis some advantage over the Warsaw basis as far as the Higgs physics is concerned.

Now we present all the operators upto dimension six in our basis of 2HDMEFT, which is motivated by the SILH basis of SMEFT. After including these operators the total Lagrangian looks like:
\begin{equation}
\mathcal{L} = \mathcal{L}_{2HDM}^{(4)} + \mathcal{L}^{(5)} + \mathcal{L}^{(6)},
\end{equation}
where, $\mathcal{L}^{(5)}$  consists of three operators, $\mathcal{O}^{(5)}_{ij} = (\tilde{\varphi_{i}}^{\dagger}\,l)^{T} C (\tilde{\varphi_{j}}^{\dagger}\,l)$ with $i,j = 1,2$, and,
\begin{eqnarray}
\mathcal{L}^{(6)} &=& \mathcal{L}_{\varphi^4 D^2} + \mathcal{L}_{\varphi^2 D^2 X} + \mathcal{L}_{\varphi^2 X^2} + \mathcal{L}_{\varphi^6} + \mathcal{L}_{\varphi^3 \psi^2} + \mathcal{L}_{\varphi^2 \psi^2 D} + \mathcal{L}_{\varphi \psi^2 X} + \mathcal{L}_{D^2 X^2} + \mathcal{L}_{\psi^4}.\nn\\
\end{eqnarray}

We have defined our notation as follows: $\varphi$, $\psi$ and $X$ stand for the two scalar doublets, fermions and gauge field strength tensors respectively. $D$ stands for a derivative. Throughout this paper, we have worked under the definition of $\mathcal{L} \supset c_{i} ( O_{i} /\Lambda^2 )$, which means all the Wilson coefficients are named according to the suffix of the corresponding operator. For example, $c_{Bij}$ is the Wilson coefficient of $O_{Bij}$. We have incorporated the $\Z_2$-violating operators along with the $\Z_2$-conserving ones, which was not the case for Ref.~\cite{Crivellin:2016ihg}. So the total number of operators in our basis is more than that of Ref.~\cite{Crivellin:2016ihg}. We have marked the $\Z_2$-violating operators in blue colour. The suppressions of these operators in a SILH scenario are given in Appendix~\ref{appendix1}.

\noindent$\bullet$ $\underline{\varphi^4 D^2}$ \\
The $\varphi^4 D^2$ operators in our basis are given in Table~\ref{table:table1}. The operators $O_{(1)21(2)}$, $O_{(1)12(2)}$, $O_{(1)22(1)}$, $O_{(2)11(2)}$ are common to both our basis and the basis introduced in Ref.~\cite{Crivellin:2016ihg}. $O_{H1H12}$, $O_{H2H12}$, $O_{T4}$ and $O_{T5}$ will not appear in the basis if one demands the $\Z_2$ symmetry to be conserved in the 6-dim Lagrangian, which is the case for Ref.~\cite{Crivellin:2016ihg}. In absence of these two operators, the number of operator in our basis is 11 compared to 12 of Ref.~\cite{Crivellin:2016ihg}. We will keep these $\Z_2$-violating operators following the logic of Section~\ref{2hdm}.
\begin{table}[h!]
\begin{center}
%\begin{table}
\begin{tabular}{|c|c|c|}
            \cline{1-3}
             \multicolumn{3}{|c|}{$\varphi^4 D^2$} \\
            \cline{1-3}

$O_{H1} = (\partial_{\m}|\varphi_1|^2)^2$ & $O_{T1} = (\varphi_1^{\dagger}\overset\leftrightarrow{D_{\m}} \varphi_1)^2$ & $O_{(1)21(2)} = (\varphi_1^{\dagger} D_{\m} \varphi_2)(D^{\m}\varphi_1^{\dagger} \varphi_2)$\\
$O_{H2} = (\partial_{\m}|\varphi_2|^2)^2$ &  $O_{T2} = (\varphi_2^{\dagger}\overset\leftrightarrow{D_{\m}} \varphi_2)^2$ & $O_{(1)12(2)} = (\varphi_1^{\dagger} D_{\m} \varphi_1)(D^{\m}\varphi_2^{\dagger} \varphi_2)$ \\
$O_{H1H2} = \partial_{\m}|\varphi_1|^2 \partial^{\m}|\varphi_2|^2$ & $O_{T3} = (\varphi_1^{\dagger}\overset\leftrightarrow{D_{\m}} \varphi_2)^2 + h.c.$ & $O_{(1)22(1)} = (\varphi_1^{\dagger} D_{\m} \varphi_2)(D^{\m}\varphi_2^{\dagger} \varphi_1)$\\
 $O_{H12} = (\partial_{\m}(\varphi_1^{\dagger} \varphi_2 + h.c.))^2$  & \textcolor{blue}{$O_{T4}=(\varphi_1^{\dagger}\overset\leftrightarrow{D_{\m}} \varphi_2)(\varphi_1^{\dagger}\overset\leftrightarrow{D_{\m}} \varphi_1)+h.c.$} & $O_{(2)11(2)} = (\varphi_2^{\dagger} D_{\m} \varphi_1)(D^{\m}\varphi_1^{\dagger} \varphi_2)$ \\
  \textcolor{blue}{$O_{H1H12} = \partial_{\m}|\varphi_1|^2\partial^{\m}(\varphi_1^{\dagger} \varphi_2 + h.c.)$}  & \textcolor{blue}{$O_{T5}=(\varphi_1^{\dagger}\overset\leftrightarrow{D_{\m}} \varphi_2)(\varphi_2^{\dagger}\overset\leftrightarrow{D_{\m}} \varphi_2)+h.c.$} &  \\
\textcolor{blue}{$O_{H2H12} = \partial_{\m}|\varphi_2|^2\partial^{\m}(\varphi_1^{\dagger} \varphi_2 + h.c.)$}  &  & \\
\hline
\end{tabular}
\caption{Operators in $\mathcal{L}_{\varphi^4 D^2}$.}
%\end{table}
\label{table:table1}
\end{center}
\end{table}
The transformations from the basis of Ref.~\cite{Crivellin:2016ihg} to our basis are:
\begin{eqnarray}
\label{basis_exchange}
O_{H1} &=& \textbf{T} - Q_{\DAlambert D}^{(1)1}, \nn\\
O_{H2} &=& \textbf{T} - Q_{\DAlambert D}^{(2)2}, \nn\\
O_{H12} &=& \textbf{T} + \textbf{E} + 2 (Q_{\varphi D}^{12(12)} + Q_{\varphi D}^{12(21)}),\nn\\
O_{H1H2} &=& \textbf{T} - Q_{\DAlambert D}^{(1)2} = \textbf{T} - Q_{\DAlambert D}^{(2)1}, \nn\\
O_{T{i}} &=& \textbf{T} + \textbf{E} + O_{H_{i}} - 4 Q^{(i)ii{i}}_{\varphi D}, \nn\\
O_{T3} &=& \textbf{T} + \textbf{E} -4 O_{(1)21(2)} - 2 Q_{\varphi D}^{12(12)},
\end{eqnarray}
where $i = 1,2$ and $\textbf{T}$ denotes total derivative terms containing $\varphi$ and $D$ and $\textbf{E}$ stands for the $\varphi^4$, $\varphi^6$ and $\varphi^3\psi^2$ terms which are already included in the basis. In the above, the fourth transformations in eqn.~(\ref{basis_exchange}) points to the fact that there is one redundant operator in the basis of Ref.~\cite{Crivellin:2016ihg}. 

\begin{table}[h!]
\begin{center}
%\begin{table}
\begin{tabular}{|c|c|}
\hline
$(D\,\varphi)(D\,\varphi)X$ & $(\varphi\,D\,\varphi) (D\,X)$ \\
\hline
$O_{\varphi B11} = ig^{\prime}(D_{\mu} \varphi_{1}^{\dagger} D_{\nu} \varphi_{1} ) B^{\m\n}$ & $O_{B11} = \frac{ig^{\prime}}{2}(\varphi_1^{\dagger} \overset\leftrightarrow{D_{\m}} \varphi_1 ) D_{\n} B^{\m \n} $  \\
 $O_{\varphi B22} = ig^{\prime}(D_{\mu} \varphi_{2}^{\dagger} D_{\nu} \varphi_{2}) B^{\m\n}$ & $O_{B22} = \frac{ig^{\prime}}{2}(\varphi_2^{\dagger} \overset\leftrightarrow{D_{\m}} \varphi_2 ) D_{\n} B^{\m \n}$  \\
 \textcolor{blue}{$O_{\varphi B12} = ig^{\prime}(D_{\mu} \varphi_{1}^{\dagger} D_{\nu} \varphi_{2}) B^{\m\n} + h.c.$}  & \textcolor{blue}{$O_{B12} = \frac{ig^{\prime}}{2}(\varphi_1^{\dagger} \overset\leftrightarrow{D_{\m}} \varphi_2 ) D_{\n} B^{\m \n}  + h.c. $} \\
 $O_{\varphi W11} = ig (D_{\mu} \varphi_{1}^{\dagger} \vec{\sigma} D_{\nu} \varphi_{1}) \vec{W}^{\m\n}$  &  $O_{W11} = \frac{ig}{2}(\varphi_1^{\dagger} \vec{\sigma} \overset\leftrightarrow{D_{\m}} \varphi_1 ) D_{\n}\vec{W}^{\m \n}  $\\
 $O_{\varphi W22} = ig (D_{\mu} \varphi_{2}^{\dagger} \vec{\sigma} D_{\nu} \varphi_{2}) \vec{W}^{\m\n}$ & $O_{W22} = \frac{ig}{2}(\varphi_2^{\dagger} \vec{\sigma} \overset\leftrightarrow{D_{\m}} \varphi_2  )D_{\n} \vec{W}^{\m \n}  $ \\
  \textcolor{blue}{$O_{\varphi W12} = ig (D_{\mu} \varphi_{1}^{\dagger} \vec{\sigma} D_{\nu} \varphi_{2}) \vec{W}^{\m\n}+ h.c.$}   &  \textcolor{blue}{$O_{W12} =  \frac{ig}{2}(\varphi_1^{\dagger} \vec{\sigma} \overset\leftrightarrow{D_{\m}} \varphi_2 ) D_{\n}\vec{W}^{\m \n}  + h.c.$   } \\
\hline
\end{tabular}
\caption{Operators in $\mathcal{L}_{\varphi^2 D^2 X}$. }
%\end{table}
\label{table2}
\end{center}
\end{table}
\noindent$\bullet$ $\underline{\varphi^2 D^2 X}$\\
We have included 12 operators of class $\varphi^2 D^2 X$, which were not there in Ref.~\cite{Crivellin:2016ihg} and are listed in Table~\ref{table2}. We have traded 6 operators of the class $\varphi^2 D^2 X$ for 6 operators of class $\varphi^2 X^2$, according to the relations:
\begin{eqnarray}
O_{Bij}  &=& \textbf{T} + O_{\varphi Bij} + \frac{1}{4}(O_{WBij} + O_{BBij}) \nn\\
O_{Wij}  &=& \textbf{T} + O_{\varphi Wij} + \frac{1}{4}(O_{WBij} + O_{WWij})
\end{eqnarray}  
where, $\textbf{T}$ stands for total derivative terms and,
\begin{eqnarray}
\label{definitns}
O_{Bij} &=& \frac{ig^{\prime}}{2} (\varphi^{\dagger}_i \overset\leftrightarrow{D_{\m}} \varphi_j) D_{\nu} B^{\m\n} \nn\\
O_{Wij} &=& \frac{ig}{2} (\varphi^{\dagger}_i \vec{\sigma} \overset\leftrightarrow{D_{\m}} \varphi_j) D_{\nu} \vec{W}^{\m\n} \nn\\
O_{\varphi Bij} &=& ig^{\prime}  (D_{\mu} \varphi_{i}^{\dagger} D_{\nu} \varphi_{j}) B^{\m\n}\nn\\
O_{\varphi Wij} &=& ig  (D_{\mu} \varphi_{i}^{\dagger} \vec{\sigma} D_{\nu} \varphi_{j}) \vec{W}^{\m\n}\nn\\
O_{VVij} &=&  g_V^2 (\varphi^{\dagger}_{i} \varphi_{j})V_{\m\n}V^{\m\n}
\end{eqnarray}
In eqn.~(\ref{definitns}), $g_{V} = g, g^{\prime}$ for $V = W^{i},B$ respectively. In our basis only $O_{BBij}$ and $O_{GGij}$ remain from class $\varphi^2 X^2$ while we have traded away $O_{WBij}$ and $O_{WWij}$ in favour of $O_{\varphi Bij}$ and $O_{\varphi Wij}$ respectively. \\

Six operators from the class $\varphi^2 \psi^2 D$ can be traded for $O_{Bij}$ and $O_{Wij}$ using, 
\bea
\label{ops}
(\varphi^{\dagger}_m \tau^{I} \overset\leftrightarrow{D_{\m}} \varphi_n) D_{\n}W^{I\m\n} &=& \sum_{i=1,2} \frac{g}{2} (\varphi^{\dagger}_m \tau^{I} \overset\leftrightarrow{D_{\m}} \varphi_n)(\varphi^{\dagger}_i \tau^{I} \overset\leftrightarrow{D^{\m}} \varphi_i) + \frac{g}{2} (\varphi^{\dagger}_m \tau^{I} \overset\leftrightarrow{D_{\m}} \varphi_n)(\bar{l} \tau^{I} \gamma^{\m}l) \nn\\
&&+ \frac{g}{2} (\varphi^{\dagger}_m \tau^{I} \overset\leftrightarrow{D_{\m}} \varphi_n)(\bar{q} \tau^{I} \gamma^{\m}q),\nn\\
%\end{align*}
%\begin{align*}
(\varphi^{\dagger}_m \overset\leftrightarrow{D_{\m}} \varphi_n) D_{\n}B^{\m\n} &=& \sum_{i=1,2} g^{\prime} Y_{\varphi_{i}} (\varphi^{\dagger}_m \overset\leftrightarrow{D_{\m}} \varphi_n)(\varphi^{\dagger}_i \overset\leftrightarrow{D^{\m}} \varphi_i) + \sum_{\psi = q,l,u,d,e} g^{\prime} Y_{\psi} (\varphi^{\dagger}_m \overset\leftrightarrow{D_{\m}} \varphi_n)(\bar{\psi}\g^{\m}\psi).\nn\\
\eea
%\end{align*}
We have removed operators $(\tilde{\varphi_i}^{\dagger}\textit{i}\,\tau^{I} \overset\leftrightarrow{D}_{\mu}\varphi_j)(\bar{l}\tau^{I}\gamma^{\mu} l)$ and $(\tilde{\varphi_i}^{\dagger}\textit{i} \overset\leftrightarrow{D}_{\mu}\varphi_j)(\bar{l} \gamma^{\mu} l)$ in favour of $O_{Bij}$ and $O_{Wij}$, as it was done in the SILH basis of SMEFT. 

As it is mentioned in Appendix~\ref{appendix1}, in a SILH scenario, $O_{\varphi Bij}$ and $O_{\varphi Wij}$ have suppressions of $\sim 1/(4\pi f)^2$, whereas  $ O_{Bij}$ and $O_{Wij}$ will be suppressed by $\sim 1/m_{\rho}^2$. Both of these operators are of type $\varphi^2 D^2 X$, but the latter ones are current-current type of operators and can be generated by integrating out suitable resonances which are typically of mass $m_{\rho}$ and couple to both the currents at the tree level. 
\\

\noindent$\bullet$ $\underline{\varphi^2 X^2}$ \\
As it was discussed for $\varphi^2 D^2 X$, some of the operators of class $\varphi^2 X^2$ were traded in favour of the previous ones. Rest of the operators in this category are listed in Table~\ref{table:table3}.\\
\begin{table}[h!]
\begin{center}
\begin{tabular}{|c|c|}
            \cline{1-2}
             \multicolumn{2}{|c|}{$\varphi^2 X^2$} \\
            \cline{1-2}
\hline
$O_{BB11} = g^{\prime 2} (\varphi_{1}^{\dagger} \varphi_{1})B_{\m\n}B^{\m\n}$ & $O_{GG11} = g_{s}^2 (\varphi_{1}^{\dagger} \varphi_{1})G^{a}_{\m\n}G^{a\m\n}$  \\
$O_{BB22} = g^{\prime 2} (\varphi_{2}^{\dagger} \varphi_{2})B_{\m\n}B^{\m\n}$ &  $O_{GG22} = g_{s}^2(\varphi_{2}^{\dagger} \varphi_{2})G^{a}_{\m\n}G^{a\m\n}$   \\
\textcolor{blue}{$O_{BB12} = g^{\prime 2} (\varphi_{1}^{\dagger} \varphi_{2} + h.c.)B_{\m\n}B^{\m\n}$} &  \textcolor{blue}{$O_{GG12} = g_{s}^2(\varphi_{1}^{\dagger} \varphi_{2} + h.c.)G^{a}_{\m\n}G^{a\m\n}$}\\
\hline
\end{tabular}
\caption{Operators in $\mathcal{L}_{\varphi^2 X^2}$.}
\label{table:table3}
\end{center} 
\end{table}

\noindent$\bullet$ $\underline{\varphi^6}$ \\
These are the corrections to the potential of the renormalizable 2HDM and are listed in Appendix~\ref{appendix3} along with the modified minimisation conditions of the potential.\\

\noindent$\bullet$ $\underline{\varphi^3 \psi^2}$ \\
These operators lead to the corrections to the Yukawa terms. It is worth noting that we have written all the possible operators without considering any $\Z_{2}$ charges of either the scalar doublets or the SM fermions. While working in a particular case of either Type I, II, X or Y 2HDM, certain operators of this category have to be put to zero depending on the discrete charges of the scalars and fermions. All the operators of this category are listed in Table~\ref{table:table4}.\\
\begin{table}[h!]
\begin{center}
\begin{tabular}{|c|c|c|}
            \cline{1-3}
             \multicolumn{3}{|c|}{$\varphi^3 \psi^2$} \\
            \cline{1-3}

\hline
$O_{e\varphi}^{111} = (\bar{l}e\varphi_1) \varphi_1^{\dagger}\varphi_1 $ & $O_{d\varphi}^{111} =(\bar{q}d\varphi_1)\varphi_1^{\dagger}\varphi_1$ & $O_{u\varphi}^{111} = (\bar{q}u\tilde{\varphi}_1)\varphi_1^{\dagger}\varphi_1$ \\
$O_{e\varphi}^{122} = (\bar{l}e\varphi_1) \varphi_2^{\dagger}\varphi_2 $ & $O_{d\varphi}^{122} =(\bar{q}d\varphi_1)\varphi_2^{\dagger}\varphi_2$ & $O_{u\varphi}^{122} =(\bar{q}u\tilde{\varphi}_1)\varphi_2^{\dagger}\varphi_2$  \\
\textcolor{blue}{$O_{e\varphi}^{112} = (\bar{l}e\varphi_1)(\varphi_1^{\dagger}\varphi_2 + h.c.)$} & \textcolor{blue}{$O_{d\varphi}^{112} = (\bar{q}d\varphi_1)(\varphi_1^{\dagger}\varphi_2 + h.c.)$} & \textcolor{blue}{$O_{u\varphi}^{112} = (\bar{q}u\tilde{\varphi}_1)(\varphi_1^{\dagger}\varphi_2 + h.c.)$}\\
\textcolor{blue}{$O_{e\varphi}^{211} = (\bar{l}e\varphi_2) \varphi_1^{\dagger}\varphi_1 $} & \textcolor{blue}{$O_{d\varphi}^{211} = (\bar{q}d\varphi_2)\varphi_1^{\dagger}\varphi_1$} & \textcolor{blue}{$O_{u\varphi}^{211} = (\bar{q}u\tilde{\varphi}_2)\varphi_1^{\dagger}\varphi_1$} \\
\textcolor{blue}{$O_{e\varphi}^{222} = (\bar{l}e\varphi_2) \varphi_2^{\dagger}\varphi_2 $} & \textcolor{blue}{$O_{d\varphi}^{222} = (\bar{q}d\varphi_2)\varphi_2^{\dagger}\varphi_2$} & \textcolor{blue}{$O_{u\varphi}^{222} = (\bar{q}u\tilde{\varphi}_2)\varphi_2^{\dagger}\varphi_2$} \\
$O_{e\varphi}^{212} = (\bar{l}e\varphi_2)(\varphi_1^{\dagger}\varphi_2 + h.c.)$ & $O_{d\varphi}^{212} = (\bar{q}d\varphi_2)(\varphi_1^{\dagger}\varphi_2 + h.c.)$ & $O_{u\varphi}^{212} = (\bar{q}u\tilde{\varphi}_2)(\varphi_1^{\dagger}\varphi_2 + h.c.)$ \\
\hline
\end{tabular}
\caption{Operators in $\mathcal{L}_{\varphi^3 \psi^2}$.}
\label{table:table4}
\end{center} 
\end{table}

\noindent$\bullet$ $\underline{\varphi^2 \psi^2 D}$ \\
Some operators of this category were traded away in favour of some $\varphi^2 D^2 X$ type of operators using eqn.~(\ref{ops}). The remaining operators are listed in Table~\ref{table:table5}. These operators contribute to various decay channels of the $W$ and $Z$ bosons. \\ 
\begin{table}[h!]
\begin{center}
\begin{tabular}{|c@{\hskip 5pt}|c@{\hskip 5pt}| c|}
            \cline{1-3}
             \multicolumn{3}{|c|}{ $\varphi^2 \psi^2 D$} \\
            \cline{1-3}
\hline
$O_{\varphi ud}^{11} = \textit{i} (\tilde{\varphi_1}^{\dagger}\textit{i} \overset\leftrightarrow{D}_{\mu}\varphi_1)(\bar{u}\gamma^{\mu}d )$ & $O_{\varphi u}^{11} =   (\tilde{\varphi_1}^{\dagger}\textit{i} \overset\leftrightarrow{D}_{\mu}\varphi_1)(\bar{u}\gamma^{\mu} u)$ & $O_{\varphi q}^{11(1)} =   (\tilde{\varphi_1}^{\dagger}\textit{i} \overset\leftrightarrow{D}_{\mu}\varphi_1)(\bar{q}\gamma^{\mu} q)$\\
$O_{\varphi ud}^{22} = \textit{i} (\tilde{\varphi_2}^{\dagger}\textit{i} \overset\leftrightarrow{D}_{\mu}\varphi_2)(\bar{u}\gamma^{\mu}d ) $ & $O_{\varphi u}^{22} =   (\tilde{\varphi_2}^{\dagger}\textit{i} \overset\leftrightarrow{D}_{\mu}\varphi_2)(\bar{u}\gamma^{\mu} u)$ & $O_{\varphi q}^{22(1)} =  (\tilde{\varphi_2}^{\dagger}\textit{i} \overset\leftrightarrow{D}_{\mu}\varphi_2)(\bar{q}\gamma^{\mu} q)$\\
\textcolor{blue}{$O_{\varphi ud}^{12} =  \textit{i} (\tilde{\varphi_1}^{\dagger}\textit{i} \overset\leftrightarrow{D}_{\mu}\varphi_2 )(\bar{u}\gamma^{\mu}d ) + h.c.$} & \textcolor{blue}{$O_{\varphi u}^{12} =   (\tilde{\varphi_1}^{\dagger}\textit{i} \overset\leftrightarrow{D}_{\mu}\varphi_2)(\bar{u}\gamma^{\mu} u) + h.c.$} & \textcolor{blue}{$O_{\varphi q}^{12(1)} =  (\tilde{\varphi_1}^{\dagger}\textit{i} \overset\leftrightarrow{D}_{\mu}\varphi_2)(\bar{q}\gamma^{\mu} q) + h.c.$} \\
$O_{\varphi e}^{11} =  (\tilde{\varphi_1}^{\dagger}\textit{i} \overset\leftrightarrow{D}_{\mu}\varphi_1)(\bar{e}\gamma^{\mu}e)$ & $O_{\varphi d}^{11} = (\tilde{\varphi_1}^{\dagger}\textit{i} \overset\leftrightarrow{D}_{\mu}\varphi_1)(\bar{d}\gamma^{\mu} d)$ & $O_{\varphi q}^{11(3)} = (\tilde{\varphi_1}^{\dagger}\textit{i}\, \tau^{I} \overset\leftrightarrow{D}_{\mu}\varphi_1)(\bar{q}\tau^{I}\gamma^{\mu} q)$\\
$O_{\varphi e}^{22} =  (\tilde{\varphi_2}^{\dagger}\textit{i} \overset\leftrightarrow{D}_{\mu}\varphi_2)(\bar{e}\gamma^{\mu}e)$ & $O_{\varphi d}^{22} =  (\tilde{\varphi_2}^{\dagger}\textit{i} \overset\leftrightarrow{D}_{\mu}\varphi_2)(\bar{d}\gamma^{\mu} d)$ & $O_{\varphi q}^{22(3)} =  (\tilde{\varphi_2}^{\dagger}\textit{i} \,\tau^{I} \overset\leftrightarrow{D}_{\mu}\varphi_2)(\bar{q}\tau^{I}\gamma^{\mu} q)$\\
\textcolor{blue}{$O_{\varphi e}^{12} =   (\tilde{\varphi_1}^{\dagger}\textit{i} \overset\leftrightarrow{D}_{\mu}\varphi_2)(\bar{e}\gamma^{\mu}e) + h.c.$} & \textcolor{blue}{$O_{\varphi d}^{12} =  (\tilde{\varphi_1}^{\dagger}\textit{i} \overset\leftrightarrow{D}_{\mu}\varphi_2)(\bar{d}\gamma^{\mu} d) + h.c.$} & \textcolor{blue}{$O_{\varphi q}^{12(3)} = (\tilde{\varphi_1}^{\dagger}\textit{i}\, \tau^{I} \overset\leftrightarrow{D}_{\mu}\varphi_2)(\bar{q}\tau^{I}\gamma^{\mu} q) + h.c.$}\\
\hline
\end{tabular}
\caption{Operators in $\mathcal{L}_{\varphi^2 \psi^2 D}$.}
\label{table:table5}
\end{center} 
\end{table}

\newpage
\noindent$\bullet$ $\underline{\varphi \psi^2 X}$ \\
These operators represent the dipole moment of the SM fermions under the SM gauge fields and are listed in Table~\ref{table:table6}, where $\sigma^{i}$ and $t^{a}$ stand for the Pauli matrices and Gell-Mann matrices respectively. \\ 
\begin{table}[h!]
\begin{center}
\begin{tabular}{|c|c|c|}
            \cline{1-3}
             \multicolumn{3}{|c|}{ $\varphi \psi^2 X$} \\
            \cline{1-3}
$O_{u G}^{1}= (\bar{q} \sigma_{\m\n} t^{a} u) \tilde{\varphi_1} G^{a\m\n}$ & $O_{u W}^{1}= (\bar{q} \sigma_{\m\n} \sigma^{i} u) \tilde{\varphi_1} W^{i\m\n}$ & $O_{u B}^{1}= (\bar{q} \sigma_{\m\n} u) \tilde{\varphi_1} B^{\m\n}$\\
\textcolor{blue}{$O_{u G}^{2}= (\bar{q} \sigma_{\m\n} t^{a} u) \tilde{\varphi_2} G^{a\m\n}$} & \textcolor{blue}{$O_{u W}^{2}= (\bar{q} \sigma_{\m\n} \sigma^{i} u) \tilde{\varphi_2} W^{i\m\n}$} & \textcolor{blue}{$O_{u B}^{2}= (\bar{q} \sigma_{\m\n} u) \tilde{\varphi_2} B^{\m\n}$}\\
$O_{d G}^{1}= (\bar{q} \sigma_{\m\n} t^{a} d) \varphi_1 G^{a\m\n}$ & $O_{d W}^{1}= (\bar{q} \sigma_{\m\n} \sigma^{i} d) \varphi_1 W^{i\m\n}$ & $O_{d B}^{1}= (\bar{q} \sigma_{\m\n} d) \varphi_1 B^{\m\n}$\\
\textcolor{blue}{$O_{d G}^{2}= (\bar{q} \sigma_{\m\n} t^{a} d) \varphi_2 G^{a\m\n}$} & \textcolor{blue}{$O_{d W}^{2}= (\bar{q} \sigma_{\m\n} \sigma^{i} d) \varphi_2 W^{i\m\n}$} & \textcolor{blue}{$O_{d B}^{2}= (\bar{q} \sigma_{\m\n} d) \varphi_2 B^{\m\n}$}\\
 & $O_{e W}^{1}= (\bar{l} \sigma_{\m\n} \sigma^{i} e) \varphi_1 W^{i\m\n}$ & $O_{e B}^{1}= (\bar{l} \sigma_{\m\n} e) \varphi_1 B^{\m\n}$\\
  & \textcolor{blue}{$O_{e W}^{2}= (\bar{l} \sigma_{\m\n} \sigma^{i} e) \varphi_2 W^{i\m\n}$} & \textcolor{blue}{$O_{e B}^{1}= (\bar{l} \sigma_{\m\n} e) \varphi_2 B^{\m\n}$}\\
\hline
\end{tabular}
\caption{Operators in $\mathcal{L}_{\varphi \psi^2 X}$.}
\label{table:table6}
\end{center} 
\end{table}

\noindent$\bullet$  $\underline{D^2 X^2\,\, \text{and}\,\, \psi^4}$\\
The counting in these two classes of operators do not change due to the insertion of a second scalar doublet in the theory, hence these operators in our basis are the same as SMEFT. For the $D^2 X^2$ type of operators we refer to Appendix~\ref{appendix2} and the list of $\psi^4$ operators can be found in~\cite{Grzadkowski:2010es}.

\section{Scalars in 2HDMEFT}
\label{scalars}
\subsection{Kinetic diagonalisation for scalars}
The kinetic terms for the scalars (except for the charged scalars)  will pick up non-diagonal parts when two of the $\varphi$\,s of $\varphi^4 D^2$ type of operators get vevs.
\begin{eqnarray}
\mathcal{L}_{kin} &=& \frac{1}{2} \left[
\begin{array}{c}
\partial_{\mu}\rho_1\\
\partial_{\mu}\rho_2\\
\end{array}
\right]^T 
\begin{bmatrix}
    1 + \frac{\Delta_{11\rho}}{2f^2}       & \frac{\Delta_{12\rho}}{4f^2}   \\
    \frac{\Delta_{12\rho}}{4f^2}       & 1 + \frac{\Delta_{22\rho}}{2f^2}   \\
\end{bmatrix}\left[
\begin{array}{c}
\partial^{\mu}\rho_1\\
\partial^{\mu}\rho_2\\
\end{array}
\right] 
+\frac{1}{2} \left[
\begin{array}{c}
\partial_{\mu}\eta_1\\
\partial_{\mu}\eta_2\\
\end{array}
\right]^T 
\begin{bmatrix}
    1 + \frac{\Delta_{11\eta}}{2f^2}       & \frac{\Delta_{12\eta}}{4f^2}   \\
    \frac{\Delta_{12\eta}}{4f^2}       & 1 + \frac{\Delta_{22\eta}}{2f^2}   \\
\end{bmatrix}\left[
\begin{array}{c}
\partial^{\mu}\eta_1\\
\partial^{\mu}\eta_2\\
\end{array}
\right]  \nn\\
 &&+\left[
\begin{array}{c}
\partial_{\mu}\phi_1^{\pm}\\
\partial_{\mu}\phi_2^{\pm}\\
\end{array}
\right]^T 
\begin{bmatrix}
    1        & 0  \\
    0       & 1    \\
\end{bmatrix}\left[
\begin{array}{c}
\partial^{\mu}\phi_1^{\pm}\\
\partial^{\mu}\phi_2^{\pm}\\
\end{array}
\right],
\end{eqnarray}
where,
\begin{eqnarray}
\Delta_{11\rho} &=& 4 c_{H1} v_1^2 + \Big(4 c_{H12} + 2 c_{T3} + c_{(1)22(1)}\Big) v_2^2 + 4c_{H1H12} v_1 v_2, \nn\\
\Delta_{22\rho} &=& \Big(4 c_{H12} + 2 c_{T3} + c_{(2)11(2)}\Big) v_1^2 + 4 c_{H2} v_2^2 + 4c_{H2H12} v_1 v_2, \nn\\
\Delta_{12\rho} &=& 2 c_{H1H12} v_1^2 + 2 c_{H2H12} v_2^2 + \Big(4 c_{H12} + 2 c_{H1H2} - 2 c_{T3} + c_{(1)21(2)} + c_{(1)12(2)}\Big) v_1 v_2, \nn\\
\Delta_{11\eta} &=&  - 4 c_{T1} v_1^2 + \Big(c_{(1)22(1)} - 2 c_{T3}\Big) v_2^2 -4 c_{T4} v_1 v_2, \nn\\
\Delta_{22\eta} &=&  \Big(c_{(2)11(2)} - 2 c_{T3}\Big) v_1^2 - 4 c_{T2} v_2^2 - 4 c_{T5} v_1 v_2, \nn \\
\Delta_{12\eta} &=&  - 4 c_{T4}v_1^2 - 4 c_{T5} v_2^2 + \Big(c_{(1)12(2)} - 2 c_{T3} -  c_{(1)21(2)}\Big) v_1 v_2 . 
\end{eqnarray}
One has to shift the fields in order to diagonalise the kinetic terms in the following manner: 
\begin{eqnarray}
\rho_1 &\rightarrow & \rho_1 \Big(1 - \frac{\Delta_{11\rho}}{4f^2}\Big) - \rho_2 \frac{\Delta_{12\rho}}{8f^2}, \nn\\
\rho_2 &\rightarrow & \rho_2 \Big(1 - \frac{\Delta_{22\rho}}{4f^2}\Big) - \rho_1\frac{\Delta_{12\rho}}{8f^2}, \nn \\
\eta_1 &\rightarrow & \eta_1 \Big(1 - \frac{\Delta_{11\eta}}{4f^2}\Big) - \eta_2\frac{\Delta_{12\eta}}{8f^2}, \nn\\
\eta_2 &\rightarrow & \eta_2 \Big(1 - \frac{\Delta_{22\eta}}{4f^2}\Big) - \eta_1\frac{\Delta_{12\eta}}{8f^2}, \nn\\
\phi^{\pm}_{1,2} &\rightarrow & \phi^{\pm}_{1,2}.
\end{eqnarray}

\subsection{Masses of the scalars}
Mass terms of the scalars are modified in the following manner:
\label{masses}
\begin{eqnarray}
\mathcal{L}_{mass} &=& \frac{1}{2} \left[
\begin{array}{c}
\rho_1\\
\rho_2\\
\end{array}
\right]^T 
(m^2_{\rho} + \Delta m^{2}_{\rho})\left[
\begin{array}{c}
\rho_1\\
\rho_2\\
\end{array}
\right] 
+\frac{1}{2} \left[
\begin{array}{c}
\eta_1\\
\eta_2\\
\end{array}
\right]^T 
(m^2_{\eta} + \Delta m^{2}_{\eta}) \left[
\begin{array}{c}
\eta_1\\
\eta_2\\
\end{array}
\right] \nn\\
&& + \left[
\begin{array}{c}
\phi_1^{\pm}\\
\phi_2^{\pm}\\
\end{array}
\right]^T 
(m^2_{\phi} + \Delta m^{2}_{\phi})\left[
\begin{array}{c}
\phi_1^{\pm}\\
\phi_2^{\pm}\\
\end{array}
\right],
\end{eqnarray}
$m^2_{\rho,\eta,\phi}$ stand for the mass matrices of the corresponding fields in the tree-level 2HDM, whereas the $\Delta m^2$s represent the contributions arising from 6-dim operators. The masses of the scalars get two-fold modifications in the EFT, one coming from the $\varphi^6$ operators, another coming from the shifting of the fields due to the $\varphi^4 D^2$ operators, 
\begin{equation}
\Delta m^2 = \Delta m^{2}_{\varphi^6} + \Delta m^{2}_{\varphi D}.
\end{equation} 
\begin{eqnarray}
\label{eqn3.6}
m^{2}_{\eta} + \Delta m^{2}_{\eta\varphi^6} &=& \Big(m_{12}^2 - \frac{1}{2}(2 \l_5 v_1 v_2 + \l_6 v_1^2 + \l_7 v_2^2) + \frac{C_1}{f^2}\Big)
\begin{bmatrix}
    \tan \beta       & -1   \nn\\
    -1       & \cot \beta   \nn\\
\end{bmatrix}, \nn\\
m^{2}_{\phi} + \Delta m^{2}_{\phi\varphi^6} &=& \Big(m_{12}^2 - \frac{1}{2}((\l_4 + \l_5) v_1 v_2 + \l_6 v_1^2 + \l_7 v_2^2) + \frac{C_2}{f^2}\Big)
\begin{bmatrix}
    \tan \beta       & -1   \nn\\
    -1       & \cot \beta   \nn\\
\end{bmatrix}, \\
\end{eqnarray}
where,
\begin{eqnarray}
C_1 &=& - \Big[v_1 v_2 (v_1^2 c_{(1212)1} + v_2^2 c_{(1212)2})\nn\\
&&+ v_1^2 v_2^2 \Big(\frac{1}{4} c_{(1221)12} + \frac{1}{4} c_{12(12)} + 3 c_{121212}\Big)  + \frac{ v_1^4}{4} c_{11(12)} +\frac{ v_2^4}{4} c_{22(12)}  \Big], \nn\\
C_2 &=& - \Big[\frac{v_1  v_2}{2 } \Big(v_1^2 (c_{(1212)1} + \frac{1}{2} c_{(1221)1}) + v_2^2 (c_{(1212)2} + \frac{1}{2} c_{(1221)2})\Big) \nn\\
&&+ v_1^2 v_2^2 \Big(\frac{3}{4} c_{(1221)12} + \frac{1}{4} c_{12(12)} + 3 c_{121212}\Big) + \frac{v_1^4}{4}c_{11(12)} + \frac{v_2^4}{4}c_{22(12)} \Big].
\end{eqnarray}
In order to arrive at eqns.~(\ref{eqn3.6}) one has to use the minimisation conditions of the modified potential which is given in Appendix~\ref{appendix3}. From eqns.~(\ref{eqn3.6}) it becomes evident that $\tan \beta = v_2/v_1$ diagonalises $(m^{2}_{\eta} + \Delta m^{2}_{\eta\varphi^6})$ and $(m^{2}_{\phi} + \Delta m^{2}_{\phi\varphi^6})$. Moreover in our case, $\Delta m^{2}_{\phi\varphi D} = 0$, because $\varphi^4 D^2$ type of operators do not affect the  charged scalar kinetic terms. This ensures that the charged scalar Goldstone boson remains massless in $\mathcal{O}(1/f^2)$. But  $\Delta m^{2}_{\eta\varphi D}$ is non-zero and is given by, 
\bea
\frac{1}{16 v_1 v_2 f^2}&&\Big(2 m_{12}^2 - (2 \l_5 v_1 v_2 + \l_6 v_1^2 + \l_7 v_2^2)\Big)\times\nn\\
&&\begin{bmatrix}
    2 v_2 (v_1 \Delta_{12\eta} - 2 v_2 \Delta_{11\eta})      & -(\Delta_{12\eta} v^2 - 2(\Delta_{11\eta} + \Delta_{22\eta})v_1 v_2)   \\
    -(\Delta_{12\eta} v^2 - 2(\Delta_{11\eta} + \Delta_{22\eta})v_1 v_2)       & 2 v_1 (v_2 \Delta_{12\eta} - 2 v_1 \Delta_{22\eta})  \\
\end{bmatrix}.
\eea
This matrix has null determinant, hence one of the eigenvalues of this matrix will be zero. This ensures that the Goldstone boson for $Z$ remains massless at $\mathcal{O}(1/f^2)$. But this matrix cannot be diagonalised by $\tan \beta = v_{2}/v_{1}$. The rotation which diagonalises the above mass matrix is denoted by some other rotation angle $\beta_{\eta}$, which can be expressed as: 
\begin{equation}
\tan \beta_{\eta} =  \Big[ 1 - \frac{\Delta_{11\eta} - \Delta_{22\eta}}{4f^2} + \frac{\Delta_{12\eta}}{8f^2}(\cot \beta -  \tan \beta)\Big]\tan \beta.
\end{equation}
In the limit $f \rightarrow \infty$, $\beta_{\eta} = \beta$,\ie we get back the tree-level 2HDM. The masses of the physical pseudoscalar and the charged scalar in terms of various Wilson coefficients are presented in Appendix~\ref{appendix4}.  

In a similar manner,  $\Delta m^{2}_{\rho\varphi^6}$ and $\Delta m^{2}_{\rho\varphi D}$ will both have non-zero values and the rotation needed for diagonalising $\Delta m^{2}_{\rho}$ will no longer be the same as $\alpha$ in 2HDM at the tree-level. We call the new rotation angle $\alpha^{\prime}$. The value of $\alpha^{\prime}$ can be determined from the masses of the neutral scalars after fixing the values of relevant Wilson coefficients. If the light and the heavy scalars have masses $m_h$ and  $m_H$ respectively~\cite{Carena:2013ooa},
\begin{eqnarray}
\sin \alpha^{\prime} &=& \frac{\mathcal{M}_{12\rho}^2}{\sqrt{(\mathcal{M}_{12\rho}^2)^2 + (\mathcal{M}_{11\rho}^2 - m_{h}^2)^2}}, \nn\\
 m_{H}^2 &=& \frac{\mathcal{M}_{11\rho}^2(\mathcal{M}_{11\rho}^2 - m_{h}^2) +  (\mathcal{M}_{12\rho}^2)^2}{(\mathcal{M}_{11\rho}^2 - m_{h}^2)}. 
\end{eqnarray}
The expressions for $\mathcal{M}_{11\rho}^2$ and $\mathcal{M}_{12\rho}^2$ in our case are given in Appendix~\ref{appendix4}.

\section{Constraints from Electroweak Precision Observables}
\label{EWPT}

We consider all the bosonic classes, \textit{i.e.}, $\varphi^4 D^2$, $\varphi^6$, $\varphi^2 D^2 X$, $\varphi^2 X^2$, and see which ones are constrained by EWPT. Some of the operators of class $\varphi^4 D^2$ contribute to the $T$ parameter and are constrained at per-mille level. But the rest of the operators of class $\varphi^4 D^2$ can only be constrained by demanding perturbative unitarity~\cite{Kikuta:2012tf}. The operators of class $\varphi^6$ do not contribute to the precision observables at all, though they can be constrained by demanding perturbative unitarity. The operators of class $\varphi^2 D^2 X$ are constrained at per-mille level by the precision tests, in particular, by the measurement of $S$ parameter and the anomalous triple gauge boson vertices~(TGV) which we will discuss shortly. The operators of class $\varphi^2 X^2$ do not contribute to the precision observables. 

The operators of class $D^2 X^2$ in our basis are the same as the ones in SMEFT, as they include no Higgs doublets. Operators of type $D^2 X^2$ contribute to the oblique parameters $V$, $W$, $Y$ and $Z$ and can be constrained from measurements at LEP. We shall mention the bounds for completeness.

We have not considered the EWPT constraints on the fermionic operator classes,\ie $\varphi^3 \psi^2$, $\varphi^2 \psi^2 D$, $\varphi \psi^2 X$ and $\psi^4$. However, we mention that the operators of class $\varphi^2 \psi^2 D$ can be constrained using the EWPT because they lead to various fermionic decay channels of $W$ and $Z$ bosons. The operators of type $\psi^4$ in our basis are the same as in SMEFT. These operators are bounded by the measurements of muon lifetime, four-fermion scatterings in LEP and LHC,\etc Operators of type $\varphi^3 \psi^2$ and $\varphi \psi^2 X$ do not contribute to the precision observables.  In this paper, we have considered only the tree-level effects of the operators of our basis on the precision observables.

In a SILH scenario, the coupling of the new dynamics is stronger compared to the SM ones,\ie  $g_{SM} \ll g_{\rho} \lesssim 4\pi$~\cite{Giudice:2007fh}. This prescription distinguishes the mass of lightest vector resonance of the strong sector $m_{\rho} \sim g_{\rho} f$ from its cut-off $\Lambda \sim 4 \pi f$. To show quantitatively what value $g_{\rho}$ can attain in a realistic scenario, in $SO(6)/SO(4)\times SO(2)$ composite Higgs model with the third generation quark doublet and $t_{R}$, both transforming as \textbf{4} of $SO(4)$ \cite{Mrazek:2011iu}, one finds that $g_{\rho} \sim$ 3.6 $< 4\pi$ for $m_{h} \sim$ 125~GeV. In the remaining part of this paper, we will use the SILH suppressions of the Wilson coefficients for different classes of operators. We have used the shorthand notation of $\xi_{i}$, where $i = 1,..,5$, to describe these suppressions and these are defined in Appendix~\ref{appendix2}. All the bounds we impose on the Wilson coefficients from now on, can be translated to a non-SILH scenario in the limit $\xi_i \rightarrow 1/f^2$, $f$ being the scale of new physics.  \\  
%\begin{itemize}

\noindent
$\bullet$ \,\,\,\textbf{Constraints from anomalous TGVs}\\

TGVs~\cite{Hagiwara:1986vm} involving $W$ bosons are some of the precisely measured quantities in the LEP experiment. Anomalous contribution to TGVs can be parametrised in terms of five parameters which are defined in Appendix~\ref{appendix5}. The operators from class $\varphi^2 D^2 X$, except $O_{Bij}$, contribute to these parameters in the following way: 
\begin{eqnarray}
\label{tgc}
\delta g^{1}_Z &=& \frac{1}{\cos^2 \theta_{w}}\Big[(c_{\varphi W11}\,c_{\b}^2 + c_{\varphi  W22}\,s_{\b}^2 + 2\,c_{\varphi W12}\,s_{\b}\,c_{\b})\xi_{2} \nn\\ 
&& + (c_{W11}\,c_{\b}^2 + c_{W22}\,s_{\b}^2 + 2 c_{W12}\,s_{\b}\,c_{\b})\xi_{1} \Big]m_{W}^2, \nn\\
\delta \kappa_{\gamma} &=& \Big[(c_{\varphi B11} + c_{\varphi W11})\,c_{\b}^2 + (c_{\varphi B22} + c_{\varphi W22}) \,s_{\b}^2 + 2\,(c_{\varphi B12} + c_{\varphi W12})\,s_{\b}\,c_{\b}\Big]m^2_{W}\xi_{2}, \nn\\ 
\delta \kappa_{Z} &=& \delta g^{1}_Z - \tan^2 \theta_{w} \delta \kappa_{\gamma}, \hspace{10pt} \lambda_{\gamma} = \lambda_{Z} = 0.
\end{eqnarray}
We use the bounds from two parameter fit ($\delta g^{1}_Z$ and $\delta \kappa_{\gamma}$) with $\lambda_{\g} = 0$ of anomalous TGVs at $95\%$ confidence level provided by LEP-II collaboration~\cite{LEPEWWG}:
\begin{eqnarray}
-4.6 \times 10^{-2} &\leqslant & \delta g^{1}_Z \leqslant  5.0\times10^{-2}, \nn\\
-1.1 \times 10^{-1} &\leqslant & \delta \kappa_{\gamma} \leqslant  8.4\times10^{-2}.
\end{eqnarray}
It is worth mentioning here that, $\l_{\g}$ and $\l_{Z}$ get affected by the CP-odd operator $W_{\m\n}W^{\m\rho}\widetilde{W}^{\n}_{\rho}$. As we are considering only the CP-even operators in this paper, $\l_{\g} = \l_{Z}=0$ in our basis. 

Inclusion of the Higgs signal strength data in the fit changes the bounds on anomalous TGVs. The fitted values of $\delta g^{1}_{Z}$ and $\delta \kappa_{\gamma}$ at $95\%$ confidence level~\cite{Falkowski:2015jaa} become:
\begin{eqnarray}
\label{lhctgc}
-1.9 \times 10^{-2} &\leqslant & \delta g^{1}_Z \leqslant  7.2\times10^{-3}, \nn\\
-2.8 \times 10^{-2} &\leqslant & \delta \kappa_{\gamma} \leqslant 0.312.
\end{eqnarray}
The 6-dim operators in our basis also contribute to the Higgs signal strengths. So, in order to extract the bounds on the Wilson coefficients using eqn.~(\ref{lhctgc}), extra care has to be taken.  \\

\noindent
$\bullet$\,\,\, \textbf{Constraints from oblique parameters}\\

%\item \textbf{Constraints from} $S$, $T$ \textbf{and} $U$ \textbf{parameters}\\
We define vacuum polarisation amplitude involving any two vector bosons $V_I$ and $V_J$ as $\Pi_{IJ}^{\m\n}(p) = \Pi_{IJ}(p^2)g^{\m\n} - \Delta(p^2) p^{\m}p^{\n}$, where  $\Pi_{IJ}(p^2) = [\Pi_{IJ}(0) + p^2 \Pi^{\prime}_{IJ}(0) + p^4 \Pi^{\prime\prime}_{IJ}(0) + ...]$. The 6-dim operators in our basis modify these polarisation amplitudes. Ward identity requires that $\Pi_{\gamma\gamma} (0) = \Pi_{\gamma Z} (0) = 0$, which we have verified. The oblique parameters $S$, $T$, $U$, $V$, $W$, $Y$ and $Z$, are expressed as different combinations of the polarisation amplitudes and their derivatives~\cite{Maksymyk:1993zm,Barbieri:2004qk,Peskin:1991sw}. Among these, only $S$, $T$ and $U$ can be measured using the $Z$-pole observables. The kinetic terms of $W_{\m}^{\pm}$ and $B_{\m}$ must be  normalised before calculating the oblique parameters~\cite{Barbieri:2004qk}. Due to the presence of the $\varphi^2 X^2$ type of operators, the kinetic term of $B_{\m}$ has to be canonically normalised with help of following transformation:
\bea
B_{\m} \rightarrow \Big(1 + g^{\prime 2} v^2 \xi_3 (c_{BB11} c_{\b}^2 + c_{BB22} s_{\b}^2 + 2 c_{BB12} s_{\b} c_{\b}) \Big) B_{\m}.
\eea 
However, such transformation need not be done for $W^{\pm}_{\m}$ as the corresponding operators do not exist in our basis, as it can be seen in Table~\ref{table:table3}.

The contribution of the effective operators to oblique parameter $U$ is zero. As, $U \propto (\Pi^{\prime}_{W^{+}W^{-}}(0) - \Pi^{\prime}_{33}(0))$, non-zero value of $U$ demands a source of isospin-violation in the theory. In our basis, these two polarisation amplitudes get modified by the operators $O_{Wij}$, but in an identical way, giving $U = 0$. 

However, the other two parameters $S$ and $T$ get non-zero contributions in our basis: 
\begin{eqnarray}
\label{obliq}
S &=&\frac{16 \pi  v^2}{m_{\rho}^2} \left[ (c_{W11} + c_{B11}) c_{\b}^2 + (c_{W22} + c_{B22}) s_{\b}^2 + 2 (c_{W12} + c_{B12}) s_{\b} c_{\b}\right], \nn\\
 T &=& \frac{1}{\alpha} \frac{v^2}{f^2}  \Big( c_{T1} c_{\b}^4 + c_{T2} s_{\b}^4 + 2 c_{T} s_{\b}^2 c_{\b}^2 + 2 c_{T4} c_{\b}^3 s_{\b} + 2 c_{T5} s_{\b}^3 c_{\b} \Big),
\end{eqnarray} 
where,
\bea
c_{T} &=&  c_{T3} - \frac{1}{8}(c_{(1)22(1)} + c_{(2)11(2)}) - \frac{1}{4}(c_{(1)21(2)} + c_{(1)12(2)}) .
\eea

As it was mentioned earlier, it can be seen that the operators that contribute to $S$ and $T$ belong to the classes $\varphi^2 D^2 X$ and $\varphi^4 D^2$ respectively.
Bounds at $95\%$ confidence limit on $S$ and $T$ are given by~\cite{Baak:2014ora}:
\bea
S \in [-0.12,0.15], \hspace{20pt} T \in [-0.04, 0.24]. \nn
\eea

The oblique parameters $V$, $W$, $Y$ and $Z$ can not be measured using the $Z$-pole observables. They represent second derivatives of certain polarisation amplitudes. $D^2 X^2$ type of operators affect the oblique parameters $V$, $W$, $Y$ and $Z$ in following manner~\cite{Giudice:2007fh}:
\bea
V &=& \Pi^{\prime\prime}_{W^{+}W^{-}}(0) - \Pi^{\prime\prime}_{33}(0) =0, \nn\\
W &=& \Pi^{\prime\prime}_{33}(0) = c_{2W} g^2 m_W^2 \xi_{4},\nn\\ 
Y &=& \Pi^{\prime\prime}_{BB}(0) = c_{2B} g^{\prime 2} m_W^2 \xi_{4},\nn\\ 
Z &=& \Pi^{\prime\prime}_{GG}(0) = c_{2G} g^{2}_{s} m_W^2 \xi_{4},
\eea
whereas the LEP data suggests~\cite{Barbieri:2004qk}: 
\bea
-4.7\times 10^{-3} \leqslant W \leqslant 0.7\times 10^{-3}, \nn\\
-0.7\times 10^{-3} \leqslant Y \leqslant 8.9\times 10^{-3}.
\eea
$V= 0$ for the same reason as why $U = 0$,\ie there is no source of isospin-violation. The parameter $Z$ could not be constrained from LEP because it is insensitive to the measurements of the electroweak sector.

One can find out the changes in the values of the precision observables due to the inclusion of any kind of new physics from its contributions to the oblique parameters\etc\cite{Burgess:1993vc}. For example, the change in $m_{W}$ and $\sin^2 \theta_{w}$ in terms of $S$, $T$, $U$ can be written as:
\begin{eqnarray}
\label{mws}
&&m_{W} = m_{Z}\bigr|_{SM} \Big(0.881 - (2.80\times10^{-3})S + (4.31\times 10^{-3})T + (3.25\times 10^{-3})U\Big), \nn\\
&&\sin^{2}_{*} \theta (m_{Z}^2) = 0.23149 + (3.64\times 10^{-3})S - (2.59\times 10^{-3})T .
\end{eqnarray}
Under the framework of 2HDMEFT, in eqn.~(\ref{mws}), $S$, $T$ and $U$ comprise of the contributions from effective operators as mentioned in eqn.~(\ref{obliq}) as well as the one-loop contributions from 2HDM~\cite{Haber:2010bw}. The `*' sign indicates that the one-loop contributions to oblique parameters have to be calculated under the Kennedy and Lynn star scheme of renormalisation~\cite{Kennedy:1988sn}. The list of the shifts in all precision observables in terms of the oblique parameters can be found in~\cite{Burgess:1994zp}. 

%%%%
\begin{figure}[h!]
 \begin{center}
\subfigure[]{
 \includegraphics[width=2.8in,height=2.8in, angle=0]{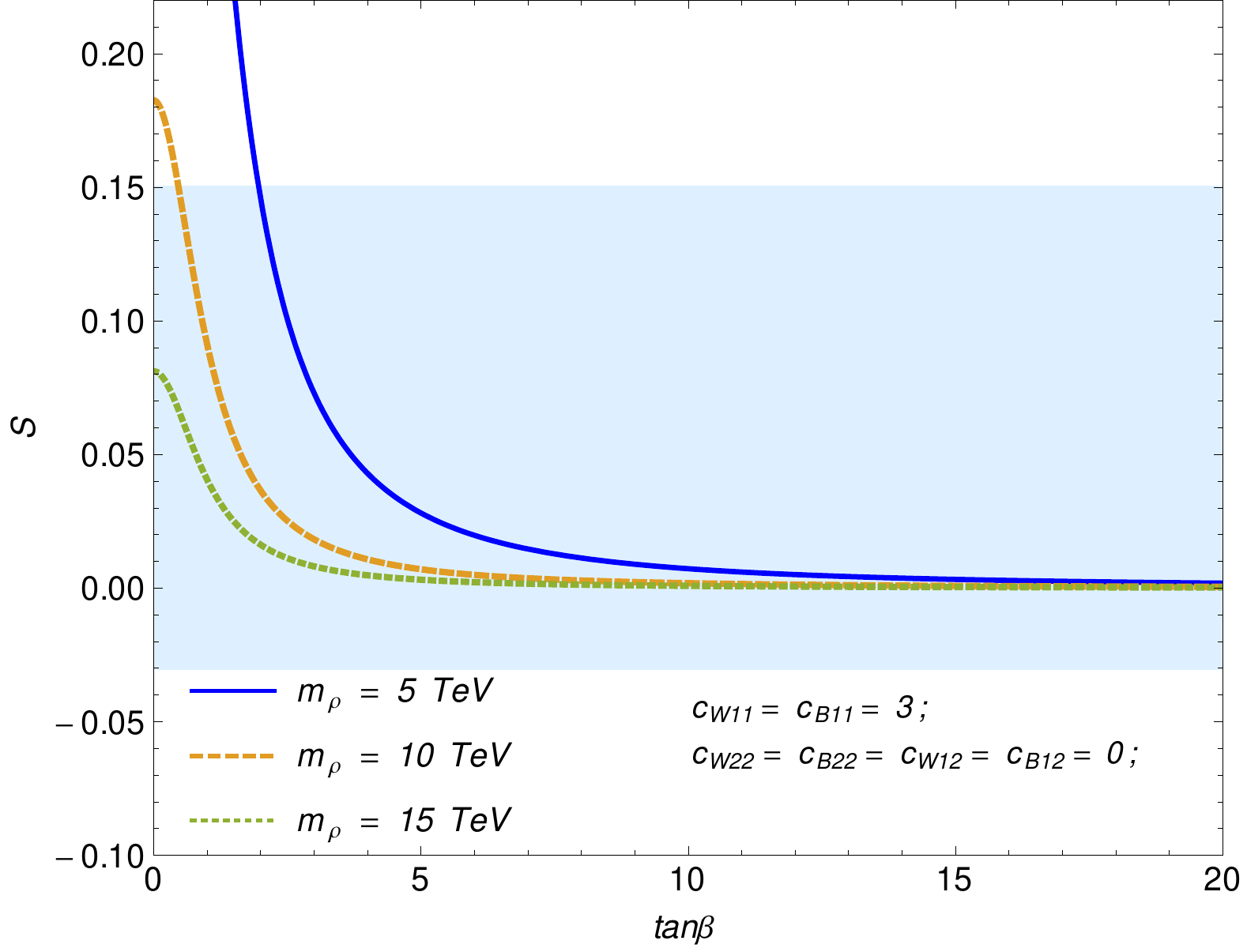}}
 \hskip 15pt
 \subfigure[]{
 \includegraphics[width=2.8in,height=2.8in, angle=0]{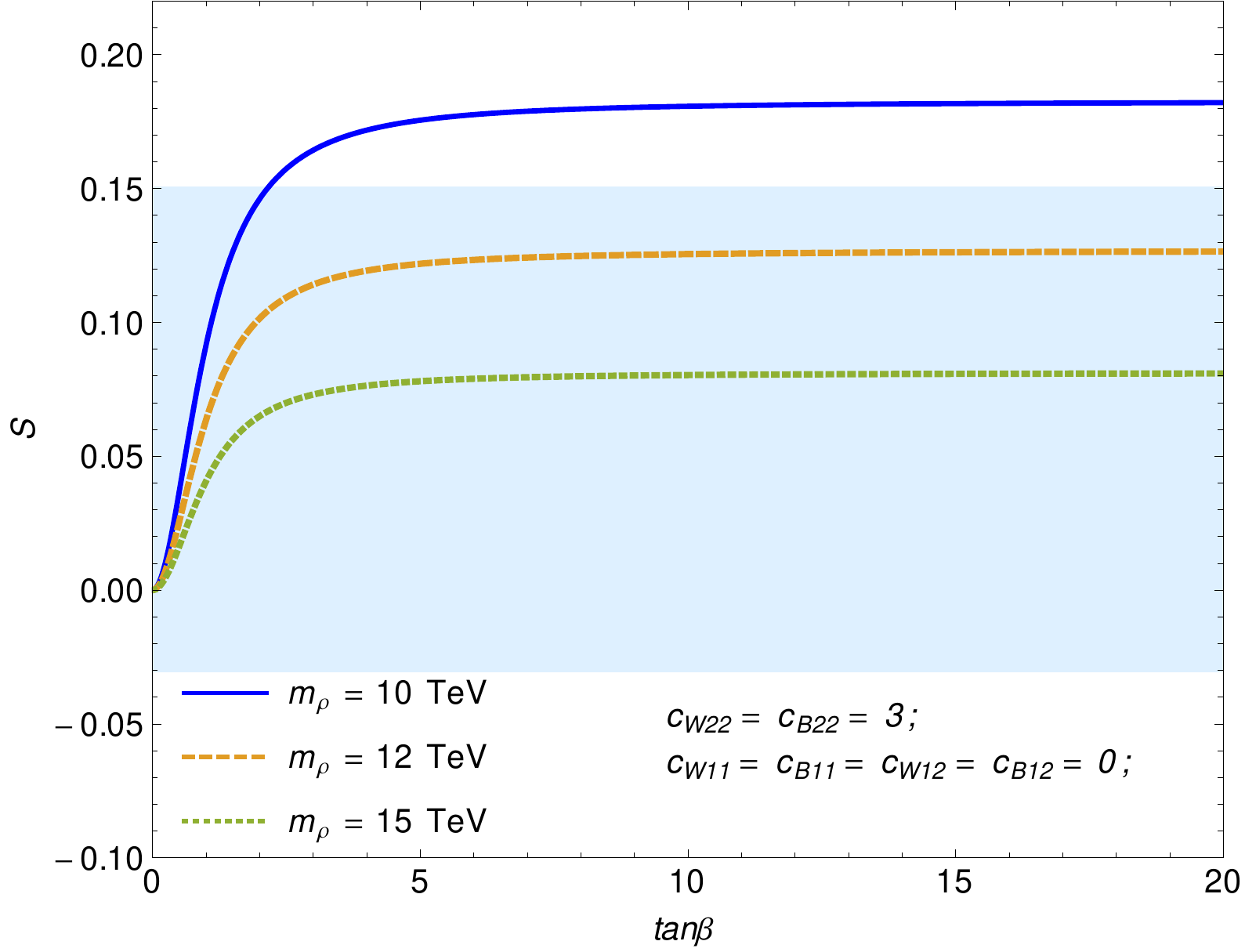}}
 \subfigure[]{
 \includegraphics[width=2.8in,height=2.8in, angle=0]{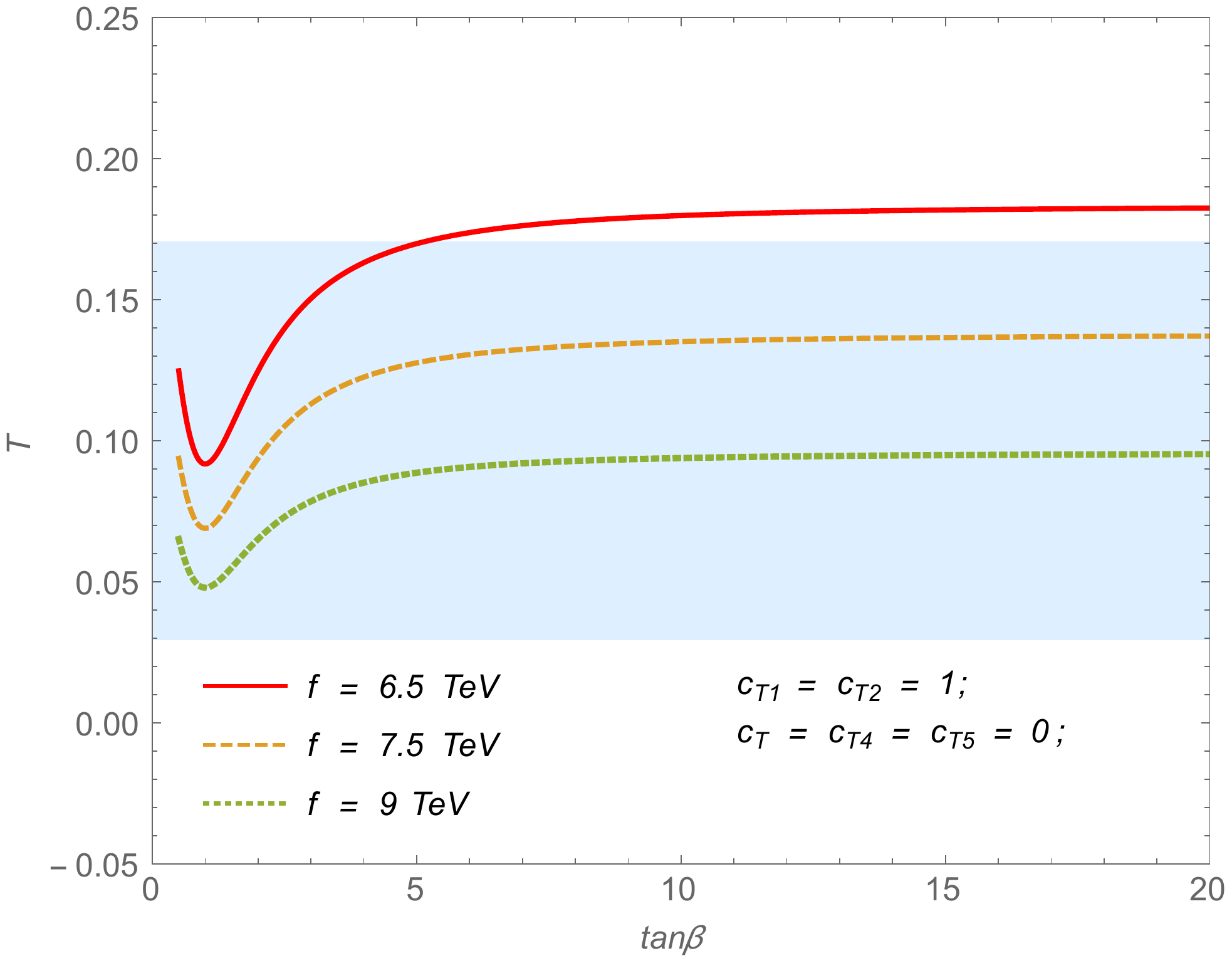}}
 \hskip 15pt
 \subfigure[]{
 \includegraphics[width=2.8in,height=2.8in, angle=0]{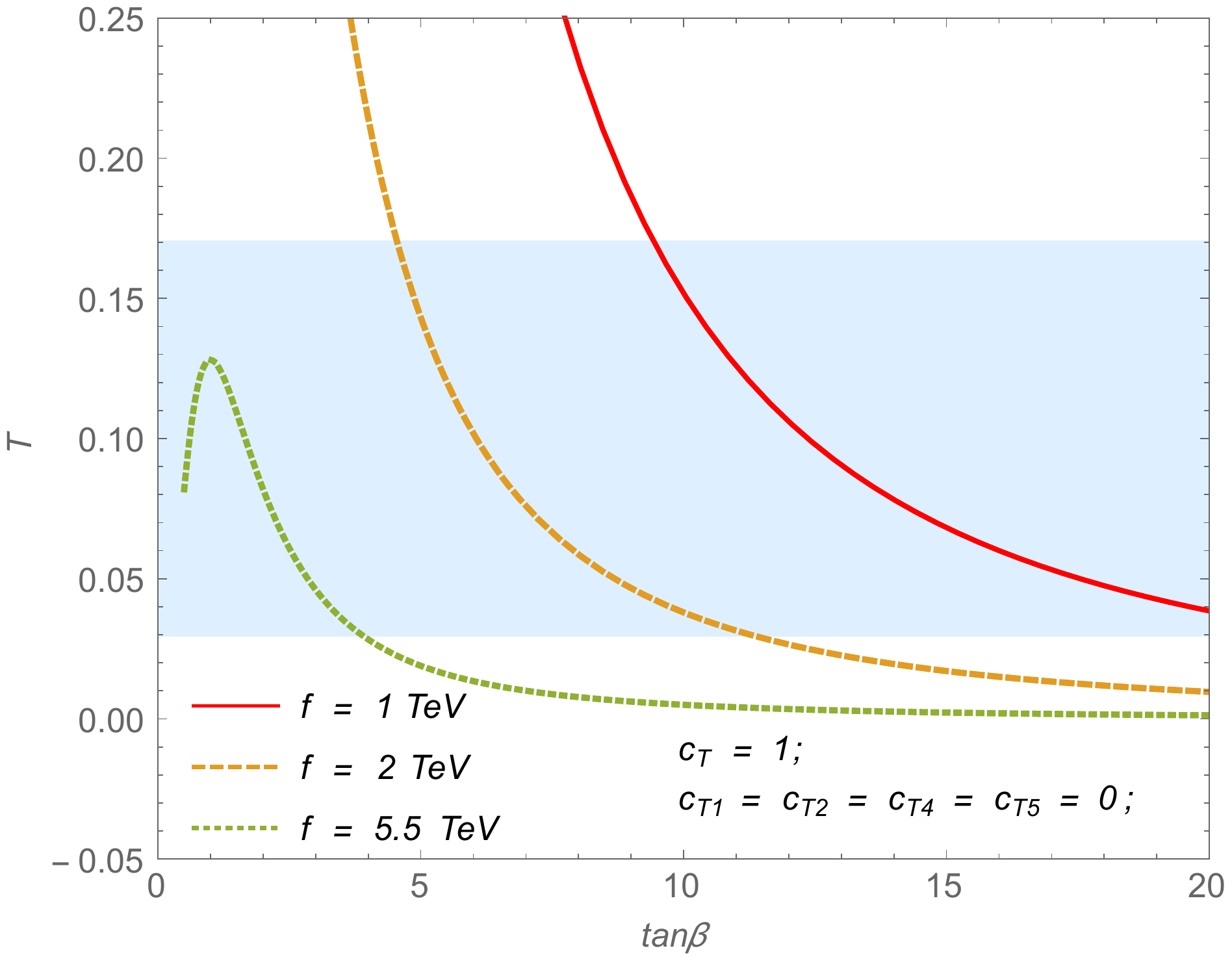}}
 \caption{ Variations of $S$ and $T$ parameter with $\tan \beta$ for different choices of Wilson coefficients. The light blue band stands for the 1$\sigma$ band for $S$ and $T$ parameters, with $U$ = 0. $S = [-0.03,0.15]$ and $T = [0.03,0.17]$~\cite{Baak:2014ora}.}
 \label{fig:STtanbeta}
\end{center}
 \end{figure}
%%%%

Based on the expressions in eqn.~(\ref{obliq}), in Fig.~\ref{fig:STtanbeta}, we have shown the dependence of $T$ and $S$ parameters on $\tan \beta$. In SMEFT one can derive the absolute value of $f$ and $m_{\rho}$ allowed by $S$ and $T$ parameters if Wilson coefficients are fixed. But in 2HDMEFT, due to the dependance of $\tan \beta$, the allowed values of $f$ and $m_{\rho}$ change significantly. In Fig.~\ref{fig:STtanbeta}(a), at lower values of $\tan \beta$, $S$ is proportonal to $c_{\beta}^2$, but as $\tan \beta \gg 1$, it becomes proportional to $(1/\tan^2 \beta)$. Fig.~\ref{fig:STtanbeta}(b) shows, at lower $\tan \beta$, $S$ varies as $s_{\beta}^2$ and becomes almost constant at higher values of $\tan \beta$. In Fig.~\ref{fig:STtanbeta}(c) and (d), $T$ varies as $(1-2s_{\beta}^2 c_{\beta}^2)$ and $2 s_{\beta}^2 c_{\beta}^2$ respectively. This is reflected as the local minima and maxima in Fig.~\ref{fig:STtanbeta}(c) and (d) respectively, around $\tan \beta = 1$. At large values of $\tan \beta$, $T$ becomes almost constant for Fig.~\ref{fig:STtanbeta}(c), whereas it becomes proportional to $1/\tan^2 \beta$ for Fig.~\ref{fig:STtanbeta}(d).

\begin{figure}[h!]
 \begin{center}
\subfigure[]{
 \includegraphics[width=2.8in,height=2.8in, angle=0]{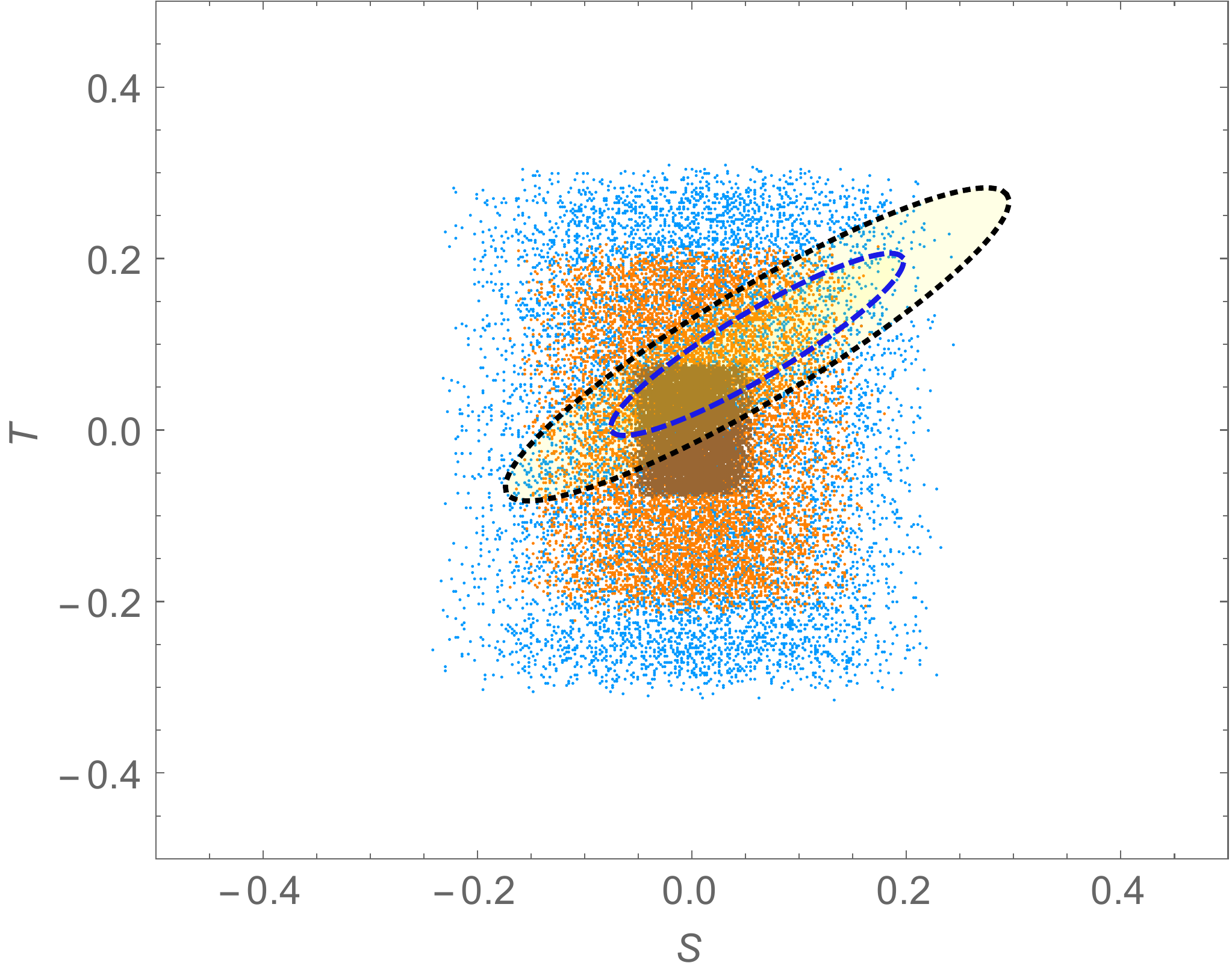}}
 \hskip 15pt
 \subfigure[]{
 \includegraphics[width=2.8in,height=2.8in, angle=0]{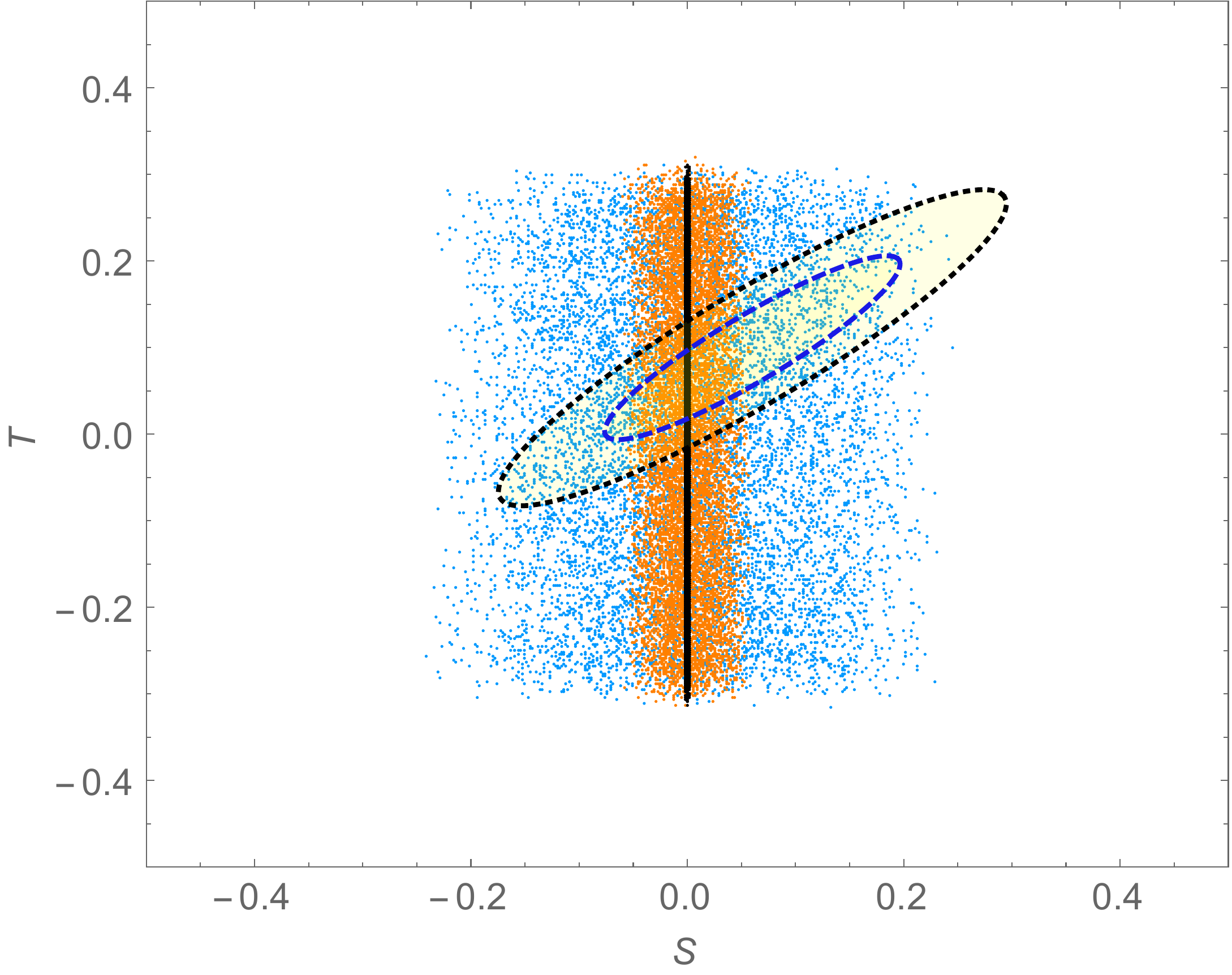}}
 \caption{Scatter plot on the $S - T$ plane with $\tan \beta = 5$ and all the Wilson coefficients taking values $[-1, 1]$. The areas enclosed by the dashed blue and dotted black lines correspond to the 1$\sigma$ and 2$\sigma$ regions respectively. (a) Blue, orange and brown points correspond to $f =$ 5, 6 and 10 TeV respectively for $g_{\rho} = 1$. (b) Blue, orange and  black points correspond to $g_{\rho} =$ 1, 2 and 10  respectively for $f = 5$ TeV.}
 \label{STdiag}
\end{center} 
 \end{figure}
In Fig.~\ref{STdiag}, we graphically illustrate the dependence of $S$ and $T$ on $f$ and $m_{\rho}$ on the $S - T$ plane. We have chosen random sets of values of the Wilson coefficients which contribute to either $S$ or $T$ for a fixed value of $\tan \beta$. In Fig.~\ref{STdiag}(a) we have seen that the higher the value of $f$, the more likely it is to satisfy the bounds of both $S$ and $T$. In Fig.~\ref{STdiag}(b), we have shown that the higher the value of $g_{\rho}$, the easier it is to satisfy the bounds from $S$ parameter. Both these effects can be read off eqn.~(\ref{obliq}). The blue points in Fig.~\ref{STdiag}(b) represent the corresponding non-SILH scenario,\ie $g_{\rho} = 1$. In Fig.~\ref{STdiag}(b), the points corresponding to $g_{\rho} = 10$ have been quenched into a straight line, indicating the fact that if the new sector is more strongly-coupled, it is easier to satisfy the bounds from $S$ parameter compared to a non-SILH scenario for the same set of values of the relevant Wilson coefficients. 

To properly disentangle the effect of the 6-dim operators in this context, we have not considered the contributions to $S$ and $T$ coming at one-loop of renormalisable 2HDM in either Fig.~\ref{fig:STtanbeta} or in Fig.~\ref{STdiag}.

\section{Constraints from observed Higgs decays to vector bosons}
\label{section:pheno}
Before going into the detailed discussion of modifications of Higgs signal strengths in the framework of 2HDMEFT in our basis, we take a moment to mention how different classes of operators contribute to Higgs physics. Operators of class $\varphi^4 D^2$ and $\varphi^6$ redefine the Higgs fields. $\varphi^6$ type of operators also contribute to the triple Higgs boson coupling and eventually to double Higgs production at LHC. $\varphi^2 D^2 X$ kind of operators induce various anomalous Lorentz structures in the Higgs coupling to the vector bosons. Among operators of class $\varphi^2 X^2$, $\mathcal{O}_{GGij}$ is constrained by the measurement of the production rate of the Higgs boson in the gluon fusion mode, whereas $\mathcal{O}_{BBij}$ is constrained from the measurement of $h \rightarrow \gamma\gamma, Z\gamma$. Coming to the fermionic operators, $\varphi^3 \psi^2$ operators contribute to both $h\bar{f}f$ and $hh\bar{f}f$ coupling. Hence, they can be constrained by the non-observation of double Higgs production. Operators of type $\varphi^2 \psi^2 D$ will lead to a non-zero $hV\bar{f}f$ coupling and eventually to the  associated production of the Higgs boson along with a massive gauge boson. 
 
After the LHC Run I and II, 2HDMs of type II, X and Y~\cite{Aoki:2009ha} are pushed close to the alignment limit. In this limit, the  couplings of one of the neutral scalars with a pair of vector bosons approach their SM values \cite{Eberhardt:2013uba, Haber:2015pua, Belusca-Maito:2016dqe}. However, alignment can be achieved with or without Appelquist-Carazzone decoupling~\cite{Appelquist:1974tg} of the new scalars. We are interested in the scenario of `alignment without decoupling'~\cite{Gunion:2002zf, Carena:2013ooa, Dev:2014yca}.

As the decay into $\g \g$, $Z \g$ and $gg$ channels are loop mediated processes in renormalisable 2HDM, there will be interferences of the one loop amplitudes with the ones coming due to higher dimensional operators. The modified matrix elements are given by: 
\begin{equation}
\label{amp}
|\mathcal{A}|^2 \simeq |\mathcal{A}^{\prime\; 2HDM}|^2 + 2 Re[\mathcal{A}^{2HDM *} \times \Delta \mathcal{A}],
\end{equation}
where, $\mathcal{A}^{2HDM}$ is the one-loop amplitude for the relevant process in 2HDM. $\mathcal{A}^{\prime\; 2HDM}$ is the amplitude of the corresponding process consisting of contributions from one-loop of 2HDM and the 6-dim operators of type $\varphi^4 D^2$, $\varphi^6$ and $\varphi^3 \psi^2$ whose effects are not studied here numerically.  $\Delta \mathcal{A}$ is the contribution coming from operators $O_{(W,\,B,\,\varphi W,\,\varphi B,\,\varphi^2 X^2)ij}$. In eqn.~(\ref{amp}), we have neglected the effects which are quadratic in the Wilson coefficients,\ie $\mathcal{O}((\Delta \mathcal{A})^2) \rightarrow 0$ and $(\mathcal{A}^{\prime \; 2HDM *} \Delta \mathcal{A}) \rightarrow (\mathcal{A}^{2HDM *}\Delta \mathcal{A})$.
After parametrising the effects of the 6-dim operators as: 
\bea
\label{coup}
\mathcal{L} \supset \Big(\frac{c_{\g\g}}{2} F_{\m\n}F^{\m\n} + c_{Z\g} Z_{\m\n} F^{\m\n} + \frac{c_{gg}}{2} G_{\m\n}G^{\m\n} \Big) \frac{h}{v}\,,
\eea
one obtains the partial decay width of SM-like Higgs to $\g\g$ and $Z\g$ with help of the eqn.~(\ref{amp})~\cite{Contino:2013kra},
\begin{eqnarray}
\Gamma (h \rightarrow \g \g)\bigr|_{EFT} &\simeq & \frac{G_{F} \a_{em}^2 m_{h}^3}{128 \sqrt{2} \pi^3} \Big[\;\bigl|\mathcal{A}^{\prime \; 2HDM}(\gamma\gamma)\bigr|^2 + 2 Re \Big[ \frac{4 \pi}{\a_{em}}\, c_{\g\g}\, \mathcal{A}^{2HDM *}(\g\g) \Big] \Big], \nn\\
\Gamma (h \rightarrow Z\g)\bigr|_{EFT} &\simeq & \frac{G_{F}^2 \a_{em} m_{W}^2 m_{h}^3}{64 \pi^4} \Big(1 - \frac{m_{Z}^2}{m_{h}^2}\Big)^3 \Big[\bigl|\mathcal{A}^{\prime \; 2HDM}(Z\g)  \bigr|^2 \nn\\
&&+ 2 Re\Big[-\frac{4 \pi}{\sqrt{\a_{em} \a_2}} c_{Z\g} \mathcal{A}^{2HDM *}(Z \g) \Big]\Big].
\end{eqnarray}
The amplitudes and relevant Wilson coefficients in our case are given as,
\begin{eqnarray}
\mathcal{A}^{(\prime)\;2HDM}(\g\g) &=& \frac{4}{3}g_{htt}^{(\prime)}\mathcal{A}^{h}_{1/2}(\tau_t) +  \frac{1}{3}g_{hbb}^{(\prime)}\mathcal{A}^{h}_{1/2}(\tau_b) + g_{h\tau\tau}^{(\prime)}\mathcal{A}^{h}_{1/2}(\tau_\tau) + \sin (\beta - \a^{(\prime)}) \mathcal{A}^{h}_{1}(\tau_W) \nn\\
&&+ \frac{m_{W}^2 \lambda^{(\prime)}_{hH^{+}H^{-}}}{2 \cos^2 \theta_{w} m_{H^{\pm}}^2} \mathcal{A}^{h}_{0}(\tau_{H^{\pm}}), \nn\\
\mathcal{A}^{(\prime) \; 2HDM}(Z\g) &=& \frac{2 \hat{v}_{t}}{\cos \theta_{w}} g_{htt}^{(\prime)}\mathcal{A}_{1/2}^{h}(\tau_t,\lambda_t) - \frac{\hat{v}_{b}}{\cos \theta_{w}}g_{hbb}^{(\prime)}\mathcal{A}^{h}_{1/2}(\tau_b,\l_b) - \frac{\hat{v}_{\tau}}{\cos \theta_{w}}g_{h\tau\tau}^{(\prime)}\mathcal{A}^{h}_{1/2}(\tau_\tau,\l_\tau) \nn\\
&&+ \sin (\beta - \a^{(\prime)}) \mathcal{A}^{h}_{1}(\tau_W,\l_W) + \frac{m_W^2 v_{H^{\pm}}}{2 \cos \theta_{w} m^2_{H^{\pm}}} \l_{h H^{+}H^{-}}^{(\prime)} \mathcal{A}^{h}_{0}(\tau_{H^{\pm}},\l_{H^{\pm}}),\nn\\
c_{\g\g} &=& 8 \sin^2 \theta_{w}( - c_{BB11} c_{\b} s_{\a} +  c_{BB22} s_{\b} c_{\a} + c_{BB12} c_{\b - \a})m_{W}^2 \xi_{3}, \nn\\
c_{Z\g} &=& \Big[( - c_{\varphi W11} c_{\b} s_{\a} +  c_{\varphi W22} s_{\b} c_{\a} + c_{\varphi W12} c_{\b - \a})\xi_{2}  \nn\\
&&- ( - c_{\varphi B11} c_{\b} s_{\a} +  c_{\varphi B22} s_{\b} c_{\a} + c_{\varphi B12} c_{\b - \a})\xi_{2} \nn\\
&&- 8( - c_{BB11} c_{\b} s_{\a} +  c_{BB22} s_{\b} c_{\a} + c_{BB12} c_{\b - \a})\xi_{3}  \Big] \tan \theta_{w} m^2_{W}, \nn\\
c_{gg} &=& 8 \Big(\frac{g_s^2}{g^2}\Big) (-c_{GG11}c_{\b}s_{\a}+ c_{GG22}s_{\b}c_{\a}+c_{GG12}c_{\b-\a})m_{W}^2\xi_{5}. 
\end{eqnarray}
In the above expressions $g_{hff}$ and $\l_{hH^{+}H^{-}}$ stand for the coupling of SM Higgs to fermion $f$ and the charged scalars $H^{\pm}$ respectively at tree level of 2HDM. Primed versions of all the couplings correspond to their values when the effects of the operators of type $\varphi^4 D^2$, $\varphi^6$ and $\varphi^3 \psi^2$ are also considered along with tree-level 2HDM. The definition of $\hat{v}_{f}$, $v_{H^{\pm}}$, the variables $\tau_{X}$ and $\l_{X}$, where, $X = t, b, \tau, W, H^{+})$ and the loop functions $\mathcal{A}^{h}_{0,1,1/2}$ can be found in~\cite{Djouadi:2005gj}. The operators $O_{BBij}$ only affect the $\g\g$ and $ZZ$ decay channels of the SM-like Higgs. On top of TGVs and the oblique parameters, the signal strengths of Higgs decaying into $Z\g$  give one further constraint on operators of type $\varphi^2 D^2 X$. The bounds on $c_{\g\g}$, $c_{Z \g}$ and $c_{gg}$ at $95\%$ CL~\cite{Falkowski:2013dza,Elias-Miro:2013mua} can be translated in our case as, 
\bea
\label{gamgam}
-0.0013 &\lesssim & c_{\g\g} \lesssim 0.0018, \nn\\
-0.016  &\lesssim & \Big[ (- c_{\varphi W11} c_{\b} s_{\a} +  c_{\varphi W22} s_{\b} c_{\a} + c_{\varphi W12} c_{\b - \a})\nn\\
&&-( - c_{\varphi B11} c_{\b} s_{\a} +  c_{\varphi B22} s_{\b} c_{\a} + c_{\varphi B12} c_{\b - \a})\Big] m_{W}^2 \xi_{2} \lesssim 0.009,\nn\\ 
-0.008 &\lesssim & c_{gg} \lesssim 0.008.
\eea

It was evident from Section~\ref{masses} that in 2HDMEFT, the tree-level couplings of the scalars to $W$ and $Z$ bosons are described by three angles,\ie $\alpha^{\prime}$, $\beta$ and $\beta_{\eta}$, rather than two, which is the case for tree-level renormalisable 2HDM. The coupling of the physical neutral scalars with the vector bosons are: 
\begin{eqnarray}
g_{HZZ} &=& \cos (\beta_{\eta} - \alpha^{\prime})\; g_{hZZ}^{SM}, \hspace{5pt} g_{HWW} = \cos (\beta - \alpha^{\prime})\; g_{hWW}^{SM}, \nn\\
g_{hZZ} &=& \sin (\beta_{\eta} - \alpha^{\prime})\; g_{hZZ}^{SM}, \hspace{5pt} g_{hWW} = \sin (\beta - \alpha^{\prime})\; g_{hWW}^{SM}.
\end{eqnarray} 
 
For simplicity, from now on we only consider the change in Higgs decay width caused by operators of class $\varphi^2 D^2 X$, namely $O_{Wij}$, $O_{\varphi Wij}$, $O_{Bij}$ and $O_{\varphi Bij}$. It was mentioned earlier in this section that operators of this class lead to anomalous Lorentz structure in the $hVV$ coupling. In presence of these operators, decay width of the SM Higgs boson into the off-shell $WW$ and $ZZ$ pairs is modified as follows~\cite{Contino:2013kra,Cahn:1988ru}: 
\bea
\label{dec1}
\Gamma (h \rightarrow V^{*}V^{(*)})\bigr|_{EFT} &=& \frac{1}{\pi^2} \int_{0}^{m_h^2} \frac{dq_1^2\, \Gamma_V M_V}{(q_1^2 - M_V^2)^2 + \Gamma_V^2 M_V^2} \int_{0}^{(m_h - q_1)^2} \frac{dq_2^2 \, \Gamma_V M_V}{(q_2^2 - M_V^2)^2 + \Gamma_V^2 M_V^2}\, \Gamma (V V)\bigr|_{EFT}, \nn\\
\eea
along with,
\begin{eqnarray}
\label{dec2}
\Gamma (VV)\bigr|_{EFT} &=& \Gamma (VV) \Big[1 - 2 \Big\{ \frac{a_{VV}}{2}\Big(1 - \frac{q_1^2+q_2^2}{m_{h}^2}\Big) + a_{V \partial V} \frac{q_1^2+q_2^2}{m_{h}^2}\Big\} \nn\\
&&+ a_{VV} \frac{\lambda(q_1^2,q_2^2,m_h^2)}{\lambda(q_1^2,q_2^2,m_h^2) + 12 q_1^2 q_2^2/ m_h^4}\Big(1 - \frac{q_1^2+q_2^2}{m_{h}^2}\Big)\Big],
\end{eqnarray}
where, 
\bea
\label{dec3}
a_{VV} &=& c_{VV} \frac{m_{h}^2}{m_{V}^2}, \hspace{10pt} a_{V \partial V} = c_{V \partial V} \frac{m_{h}^2}{2 m_{V}^2}, \nn\\
\Gamma (VV) &=& \sin^2 (\b - \a)\frac{\delta_V G_{F} m_h^3}{16 \sqrt{2} \pi} \sqrt{\lambda(q_1^2,q_2^2,m_h^2)}\Big(\lambda(q_1^2,q_2^2,m_h^2) + \frac{12 q_1^2 q_2^2}{m_h^4}\Big),
\eea
with $\delta_V = 2,1$ for $V = W,Z$ respectively, and $\lambda(x,y,z) = (1 - x/z - y/z)^2 - 4xy/z^2$. Definitions of $c_{\{WW,\,ZZ,\,W\partial W,\,Z\partial Z\}}$ are given in Appendix~\ref{appendix5}. In our basis they can be written in terms of the Wilson coefficients in the following way:
\begin{eqnarray} 
\label{wcs}
c_{WW} &=& -2 c_{\varphi W}, \hspace{10pt} c_{W\partial W} = c_{WW} - 2 c_{W}, \nn\\
c_{ZZ} &=& c_{W \partial W} - (2 c_{\varphi B} - 8 c_{BB})\tan^2\theta_{w}, \hspace{10pt} c_{Z\partial Z} = c_{W \partial W} - 2 (c_{B} + c_{\varphi B})\tan^2\theta_{w}, \nn\\
c_{\varphi W} &=& ( - c_{\varphi W11} c_{\b} s_{\a} +  c_{\varphi W22} s_{\b} c_{\a} + c_{\varphi W12} c_{\b - \a}) m_{W}^2 \xi_{2}, \nn\\
c_{\varphi B} &=& ( - c_{\varphi B11} c_{\b} s_{\a} +  c_{\varphi B22} s_{\b} c_{\a} + c_{\varphi B12} c_{\b - \a}) m_{W}^2 \xi_{2}, \nn\\
c_{BB} &=& ( - c_{BB11} c_{\b} s_{\a} +  c_{BB22} s_{\b} c_{\a} + c_{BB12} c_{\b - \a}) m_{W}^2 \xi_{3},  \nn\\
c_{W} &=& (- c_{W11}c_{\b} s_{\a} + c_{W22}s_{\b} c_{\a} +  c_{W12}c_{\b - \a})m_{W}^2 \xi_{1}, \nn\\
c_{B} &=& ( - c_{B11} c_{\b} s_{\a} +  c_{B22} s_{\b} c_{\a} + c_{B12} c_{\b - \a}) m_{W}^2 \xi_{1}.
\end{eqnarray}

One can see in eqn.~(\ref{wcs}), all the Wilson coefficients are constrained by either $S$ parameter in eqn.~(\ref{obliq}) or anomalous TGVs eqn.~(\ref{tgc}) or the measurement of decay width of Higgs to $\g\g$ and $Z\g$ in eqn.~(\ref{gamgam}). This happens in SMEFT as well. But these Wilson coefficients appeared with prefactors different from those in eqns.~(\ref{tgc}) and ~(\ref{obliq}). This is a remarkable feature of 2HDMEFT. For example, the Wilson coefficient of $O_{W11}$ has come with a prefactor of $c_{\b}^2$ and $-c_{\b}s_{\a}$ in the expressions for $S$ parameter and $hWW$ coupling respectively, and would come with a prefactor of $s_{\a}^2$ in the $hhWW$ coupling. This effect is absent in SMEFT. The following numerical analysis will illustrate this fact. 

We take four benchmark points~(BP) involving different sets of Wilson coefficients to illustrate the effect of the corresponding 6-dim operators on partial decay width of $h$. Before we start, we denote,  $\tilde{c}_{k11} c_{\b}^2 + \tilde{c}_{k22} s_{\b}^2 + 2 \tilde{c}_{k12} c_{\b} s_{\b} = \widetilde{C}_{k}$, where, $k = \{W,\,B,\,\varphi W,\,\varphi B\}$ and $\tilde{c}_{kij} = c_{kij}\, \xi_{k}$, with $\xi_{k} = \xi_{1},\xi_{1},\xi_{2},\xi_{2}$, for $k = W,B, \varphi W, \varphi B$ respectively.

\begin{itemize}
\item{\textbf{BP1}} \hspace{4pt}$\widetilde{C}_{W}\approx -10^{-3}$, $\widetilde{C}_{B} \approx -2 \times 10^{-3}$, $\widetilde{C}_{\varphi W}\approx - 10^{-2}$, $\widetilde{C}_{\varphi B} \approx - 10^{-3}$, $\tan \b = 2$ and $c_{\b - \a} = 0.1$, $s_{\b - \a} \sim 0.995$, $\tilde{c}_{k ij} \approx 0.55\, \widetilde{C}_{k}$.

\item{\textbf{BP2}} \hspace{4pt} $\widetilde{C}_{W}\approx -10^{-3}$, $\widetilde{C}_{B} \approx -2 \times 10^{-3}$, $\widetilde{C}_{\varphi W}\approx - 10^{-2}$, $\widetilde{C}_{\varphi B} \approx - 10^{-3}$, $\tan \b = 1$ and $c_{\b - \a} = 0.1$, $s_{\b - \a} \sim 0.995$. Wilson coefficients for all $\Z_2$-violating operators are set to zero; $\tilde{c}_{k22} \approx 3\,\tilde{c}_{k11} \approx 1.5\,\widetilde{C}_{k}$.

\item{\textbf{BP3}} \hspace{4pt} $\widetilde{C}_{W}\approx -10^{-3}$, $\widetilde{C}_{B} \approx -2 \times 10^{-3}$, $\widetilde{C}_{\varphi W}\approx - 10^{-2}$, $\widetilde{C}_{\varphi B} \approx - 10^{-3}$, $\tan \b = 1$ and $c_{\b - \a} = 0$, $s_{\b - \a} \sim 1$, which corresponds to pure alignment limit. Moreover we set the Wilson coefficients for all $\Z_2$-violating operators to zero; $\tilde{c}_{k22} \approx 3\,\tilde{c}_{k11} \approx 1.5\,\widetilde{C}_{k}$.

\item{\textbf{BP4}} \hspace{4pt}  $\widetilde{C}_{W}\approx -10^{-3}$, $\widetilde{C}_{B} \approx -2 \times 10^{-3}$, $\widetilde{C}_{\varphi W}\approx - 10^{-2}$, $\widetilde{C}_{\varphi B} \approx - 10^{-3}$, $\tan \b = 1$ and $c_{\b - \a} = 0.1$, $s_{\b - \a} \sim 0.995$. $\tilde{c}_{k11} \approx \frac{1}{3}\tilde{c}_{k22} \approx -\frac{1}{3}\tilde{c}_{k12} \approx \widetilde{C}_{k}$.

\end{itemize}

\begin{table}[h!]
\begin{center}
\begin{tabular}{|c|c|c|c|c|c|c|}
\hline
 &$x_{WW} = x_{ZZ}$ & $y_{WW}$ & $\bigl|y_{WW}/x_{WW}\bigr|$ & $y_{ZZ}$ & $\bigl|y_{ZZ}/x_{ZZ}\bigr|$ \\
\hline
BP1 & $-1\%$ & $-2.4\%$ & $2.4$ & $-2.1\%$ & $2.1$\\
BP2 & $-1\%$ & $-4.1\%$ &  $4.1$   & $-3.6\%$ & $3.6$\\
BP3 & 0 & $-3.9\%$ &     & $-3.4\%$ & \\ 
BP4 & $-1\%$ & $-7.0\%$ & 7.0 & $-6.1\%$ & 6.1\\
\hline
\end{tabular}
\caption{Relative changes in decay width of $h \rightarrow V^{*}V^{(*)}$.}
\label{table:table7}
\end{center} 
\end{table}

In Table~\ref{table:table7}, $x_{VV} = (\Gamma_{VV,2HDM}^{tree} - \Gamma_{VV,SM}^{tree})/\Gamma_{SM}^{tree}$ and $y_{VV} = (\Gamma_{VV,EFT} - \Gamma^{tree}_{VV,SM})/\Gamma_{VV,SM}^{tree}$. Here $\Gamma_{VV,EFT}$ consists of contributions from tree-level SM and 6-dim operators of 2HDMEFT, and can be obtained from eqns.~(\ref{dec1}), (\ref{dec2}) and (\ref{dec3}) by putting $\sin^2 (\b-\a) = 1$. We have also indicated the ratios between $y_{VV}$ and $x_{VV}$ for both $WW$ and $ZZ$ decay channel. Note that in all the cases we have kept $c_{BB} \approx 0$, which is constrained at per-mille level by the measurement of $\Gamma(h \rightarrow \g\g)$. All the BPs satisfy the TGV constraints, $S$ parameter and $\Gamma (h \rightarrow \g\g, Z\g)$ without any fine-tuning between Wilson coefficients. For all four BPs, it can be seen that the effects of 6-dim operators can be substantial at the alignment limit, which also implies that the effect of 6-dim operators are large enough to confuse the bounds on $c_{\b - \a}$ derived considering the tree-level effects in 2HDM only. BP3 mimics the situation of SMEFT, because the values of $\tan \b$ and $s_{\b -\a}$ are conspired in such a way that the combination of Wilson coefficients that enters in the $S$ parameter is the same as the one appearing in $\Gamma(h\rightarrow WW)$. A comparison between BP2 and BP3 indicates that effects of 2HDMEFT can be even larger than SMEFT ones for similar values of Wilson coefficients. In BP4 we have retained the $\Z_2$-violating operators of all the four classes, and comparing with BP2, it can be seen that the inclusion of $\Z_2$-violating operators can lead to enhanced modifications in Higgs decay widths compared to the case where only $\Z_2$-conserving operators are kept.

We have refrained from considering the one-loop effects in $h \rightarrow WW, ZZ$ to disentangle the effect of EFT in these processes, as we had done in Section~\ref{EWPT}. The effects of the 6-dim operators can be of the same order of the one-loop effects in 2HDM. For example, the percentage change in the decay widths of the processes $h \rightarrow V^{*} V^{(*)} \rightarrow 4f$ at one-loop order compared to the lowest order is around $\sim 2.7\%$, for $c_{\b - \a} = 0.1$~\cite{Altenkamp:2017ldc}. 

 In this section we have seen that the operators which are constrained via $S$ parameter, TGVs and the decay width of SM like Higgs to $\g\g$ and $Z\g$, can still be exploited to impart a change on the $h \rightarrow VV$ decay widths. At the end of LHC Run II, the error in Higgs couplings are expected to decrease upto 4\%. The change in decay widths as mentioned in Table~\ref{table:table7} can be probed in HL-LHC which will be able to probe the $hVV$ couplings to a precision of 2\%. The ILC with $\sqrt{s} = 500$ GeV will reduce the error in the $hVV$ couplings upto $\sim 1\%$~\cite{Dawson:2013bba}. If 2HDM is close to the alignment limit, the effects of these 6-dim operators will be significant and will lead to a confusion with the new contributions from 2HDM at tree and loop level effects for the respective processes.

\section{Discussions} 
\label{sec:conclude}
Non-observation of any beyond standard model particle in the direct searches at LHC motivates us to adhere to the language of effective field theory. 2HDM is a viable extension of the scalar sector of the SM. The main motivation for considering a 2HDMEFT comes from the fact that observations of a SM-like Higgs boson has pushed 2HDM to be at the alignment limit if the new scalars are at a sub-TeV scale. In such a scenario, the new scalars get almost decoupled at the vertices with SM particles that include a gauge boson. As a result, deviations of the contribution of four-dimensional Lagrangian of 2HDM to SM processes are small compared to its SM counterpart. We have shown that contributions of the six-dimensional operators of 2HDMEFT are comparable with the deviation due to tree level contributions in 2HDM at the alignment limit from SM. Such effects can interfere in determination of 2HDM parameter space from experiments. 

In this work, we have presented a complete basis for six-dimensional operators in 2HDM motivated by the SILH~\cite{Giudice:2007fh} in SMEFT. Such an extension is not trivial and demands careful use of EoMs to eliminate redundant operators. For simplicity, we have restricted ourselves to CP- and flavour-conserving ones. Due to various reasons, as mentioned in the text, in SMEFT, the SILH basis is often favoured for Higgs physics studies. Hence, we feel that our basis would be useful for the community practising 2HDM phenomenology.

In presence of a strongly interacting weak sector just beyond a TeV or so, which we designate in this paper as `SILH scenario', the hierarchy problem in Higgs mass reduces to a `Little' hierarchy problem, thereby alleviating the quadratic divergences. The 2HDM can originate from such an underlying strong dynamics. In this case, one need not bother about hard $\Z_2$-violating terms and hence, we include them all, even in the six-dimensional Lagrangian. In a SILH scenario the Wilson coefficients of the operators come with various suppression factors which we review in Appendix~\ref{appendix1}. In passing, we emphasize that our basis has a wider applicability -- it is valid when such a SILH scenario is envisaged or not.

Next, we have let the operators confront the results from the electroweak precision tests and measured values of Higgs production and decay channels concentrating only on bosonic operators of classes $\varphi^4 D^2$, $\varphi^6$, $\varphi^2 D^2 X$ and $\varphi^2 X^2$ containing the new scalars. Ensuing bounds on combinations of Wilson coefficients have been extracted.  Out of these classes, some of the operators belonging only to the classes $\varphi^4 D^2$ and $\varphi^2 D^2 X$ were constrained from EWPT. The operators of class $\varphi^6$ and some of the operators of class $\varphi^4 D^2$ can be constrained demanding perturbative unitarity. As expected, constraints from EWPT are much tighter than those from considerations of unitarity. We also consider the class $D^2 X^2$  which contribute to $W$, $Y$ and $Z$ parameters in EWPT. For completeness, we have mentioned bounds on these as well, but these are the same as in SMEFT. In discussing impacts on Higgs signal strengths, only the operators of the class $\varphi^2 D^2 X$ and $\varphi^2 X^2$ have been considered for simplicity. As the phenomenology of 2HDMEFT is quite rich, to avoid cluttering of information we have also avoided considering phenomenology of the fermionic operators of classes $\varphi^3 \psi^2$, $\varphi^2 \psi^2 D$, $\varphi \psi^2 X$ and $\psi^4$. These issues will be addressed elsewhere.

An earlier attempt to present a basis of $\Z_2$-conserving operators for 2HDMEFT was made in Ref.~\cite{Crivellin:2016ihg} that resembles the Warsaw basis in SMEFT. But we have found one redundant operator in the class $\varphi^4 D^2$. In our basis, kinetic mixing of the gauge eigenstates of the scalar were taken care of by field redefinitions and their effect on determining the masses of the physical scalar were also calculated. An important feature of our basis is that the charged scalar mass matrix is still diagonalised by $\tan\beta = v_2/v_1$ as in the tree-level 2HDM. But the neutral psuedoscalar sector needs a diagonalisation matrix which is not characterised by the same $\tan\beta$. So the kinetic diagonalisation changes the pseudoscalar mass matrix, but not the one for the charged scalars. This is reflected in the expressions for masses of the pseudoscalar and the charged scalars that we have presented in Appendix~\ref{appendix4}. 

An interesting feature of 2HDMEFT  is that, in contrast to SMEFT, here the 2HDM parameter space plays a crucial role while placing bounds on the Wilson coefficients. For example, the same Wilson coefficient can appear with different prefactors in the expressions for precision parameters and for Higgs decay widths. These prefactors depend on the 2HDM parameter space. This happens because unlike SM, in 2HDM the interaction eigenstates of the scalars are not the same as their mass eigenstates. In Section~\ref{section:pheno} we have numerically demonstrated this remarkable effect which is absent in SMEFT. 

In short, our complete basis of 2HDMEFT will facilitate further studies of 2HDM phenomenology. We have presented constraints on some of the operators and pointed out that such constraints do depend on the 2HDM parameter space. Such dependence can significantly modify some of the predictions of SMEFT. It was also noticed that, in the vicinity of the alignment limit, the effects of the higher dimensional operators in determining the parameter space of 2HDM are not negligible.

%==============================================================================
\section{Acknowledgements}
S.K. thanks A. Falkowski for email communications. S.R. is supported by the Department of Science and Technology, India via Grant No. EMR/2014/001177.

\newpage
\appendix 

\section{Rules for dimensional analysis}
\label{appendix1}
If the BSM strong sector is characterized by mass-scale $m_{\rho}$ compared to $m_{W}$ for SM, the effective operators induced by the former will be made of local operators made out of SM fields and derivatives \cite{Luty:1997fk,Pomarol:2014dya}: 
\begin{equation*}
\mathcal{L}_{EFT} = \frac{m_{\rho}^4}{g^{2}_{\rho}}\,\mathcal{L}\,\Big( \frac{D_{\mu}}{m_{\rho}},
\frac{g_{\rho}\varphi}{m_{\rho}},
\frac{g_{\rho}f_{L,R}}{m_{\rho}^{3/2}},
\frac{g_{SM}X}{m_{\rho}^2}  \Big)
\end{equation*} 
$\varphi$, $f_{L,R}$ and $X$ stand for the scalars, fermions and field strengths of SM gauge fields respectively. The above relation can be described by putting $\hbar \neq 1$ and counting the dimension of all SM fields and $g_{\rho}$ in powers of $\hbar$. One finds that, $[g_{\rho}] = \hbar^{-1/2}$, $[\varphi] = [V_{\mu}] = [M] \hbar^{1/2}$\etc 
The naive dimensional counting rules for the Wilson coefficients of operators in SILH basis, as were introduced in \cite{Giudice:2007fh}, are based on above expansion,
\begin{enumerate}[ {(}I{)}]
\item A factor of $1/f$ for an extra Goldstone leg;
\item A factor of $1/m_{\rho}$ for an extra derivative,\ie $1/m_{\rho}^{2}$ for an extra $X$;
\item A suppression of $1/m_{\rho}^2$ along with an extra SM vector boson field strength.
\end{enumerate} 
Extra suppressions in an SILH scenario are as follows: $\varphi^{4} \,D^{2}$, $\varphi^{6}$, $\varphi^{2} \psi^{2}D$ and $\varphi^{3} \psi^{2}$ are suppressed by $1/f^2$, following rule (I). $\varphi^{2} X^{2}$, $\varphi^{2} D^{2} X$ and $\varphi \psi^{2} X$ should have been suppressed by $1/m_{\rho}^2$, using rule (II). But the operators of type $O_{\varphi W}$ and $O_{\varphi B}$ can not be generated at the tree level by integrating out a new resonance and thus come with a suppression of $1/(4 \pi f)^2$. Physical Higgs is neutral under $SU(3)_{C}\times U(1)_{em}$. So, the gauging of these groups do not break the shift symmetry of physical pNGB Higgs. On the other hand, operators of type $\varphi^2 X^2$ generate the coupling of physical Higgs with a pair of on-shell photons and gluons, for $X = B,G$ respectively and also break the shift symmetry. The latter fact is reflected in an extra suppression of $(g_{SM}/g_{\rho})^2$ for the $\varphi^2 X^2$ type of operators. Moreover, operator of type $\varphi^3 \psi^2$ get an extra suppression of the Yukawa coupling $y_{\psi}$. In the similar manner, in our basis of 2HDMEFT, operators of type $(\bar{l}e\varphi_{i})(\varphi_{j}^{\dagger} \varphi_{k})$, $(\bar{q}d\varphi_{i})(\varphi_{j}^{\dagger} \varphi_{k})$ and $(\bar{q}u\tilde{\varphi_{i}})(\varphi_{j}^{\dagger} \varphi_{k})$ from Table~\ref{table:table4} get suppressed by $Y^{e}_{i}$, $Y^{d}_{i}$ and $Y^{u}_{i}$ respectively.

The SILH basis of 6-dim operators in SMEFT with their corresponding suppressions are given in Appendix~\ref{appendix2}.

\section{SILH Lagrangian of SMEFT}
\label{appendix2}
\begin{eqnarray*}
\mathcal{L}^{SILH}_{SMEFT} &=& \frac{c_{H}}{2f^2}(\partial_{\m}|H|^2)^2 + \frac{c_{T}}{2f^2}(H^{\dagger}\overset\leftrightarrow{D_{\m}} H)^2 - \frac{c_{6}\l}{f^2}|H|^6 + \Big(\frac{c_{y} y_{f}}{f^2} |H|^2 \bar{f}_{L} H f_{R} + h.c. \Big) \\
&&+ \frac{i c_W g}{2 m_{\rho}^2}(H^{\dagger}\vec{\sigma}\overset\leftrightarrow{D_{\m}} H)D_{\n}\vec{W}^{\m\n} + \frac{i c_{B} g^{\prime}}{2 m_{\rho}^2} (H^{\dagger}\overset\leftrightarrow{D_{\m}} H)D_{\n}B^{\m\n} \\
&&+ \frac{i c_{HW} g}{16 \pi^2 f^2}(D^{\m} H)^{\dagger}\vec{\sigma}(D^{\n}H)\vec{W}_{\m\n} +\frac{i c_{HB} g^{\prime}}{16 \pi^2 f^2}(D^{\m} H)^{\dagger} (D^{\n} H)B_{\m\n} \\
&&+\frac{c_{\g} g^{\prime 2}}{16 \pi^2 f^2} \frac{g^2}{g_{\rho}^2} |H|^2 B_{\m\n} B^{\m\n} + 
\frac{c_{g} g_{s}^2}{16 \pi^2 f^2} \frac{y_{t}^2}{g_{\rho}^2} |H|^2 G^{a}_{\m\n} G^{a\m\n} \\
&&- \frac{c_{2W}g^2}{2g_{\rho}^2m_{\rho}^2}(D^{\m}W_{\m\n})^{i}(D_{\rho}W^{\rho\n})^{i}  
- \frac{c_{2B}g^{\prime 2}}{2g_{\rho}^2m_{\rho}^2}(\partial^{\m}B_{\m\n})(\partial_{\rho}B^{\rho\n})  - \frac{c_{2G}g_{s}^2}{2g_{\rho}^2m_{\rho}^2}(D^{\m}G_{\m\n})^{a}(D_{\rho}W^{\rho\n})^{a}.
\end{eqnarray*}
We have used shorthand notations for the suppressions of various kinds of operators:  
\bea
\xi_{1} = \frac{1}{m_{\rho}^2},\hspace{8pt}\xi_{2} = \frac{1}{(4\pi f)^2},\hspace{8pt}\xi_{3} = \frac{g^2}{g_{\rho}^2}\frac{1}{(4\pi f)^2},\hspace{8pt}\xi_{4} = \frac{1}{g_{\rho}^2 m_{\rho}^2},\hspace{8pt}\xi_{5} = \frac{y_t^2}{(4\pi f)^2}\frac{1}{g_{\rho}^2}.\nn
\eea
\section{The potential}
\label{appendix3}
The total potential in 2HDMEFT is given by $ V(\varphi_1,\varphi_2) + \mathcal{L}_{\varphi^6}$. $V(\varphi_1,\varphi_2)$ is given in eqn.~(\ref{potential}), and, 
\begin{eqnarray*}
\mathcal{L}_{\varphi^6} &=& \frac{1}{f^2}\Big[c_{111} |\varphi_1|^6 + c_{222} |\varphi_2|^6 
+ c_{112} |\varphi_1|^4 |\varphi_2|^2 + c_{122} |\varphi_1|^2 |\varphi_2|^4 \nn\\
&&+ c_{(1221)1} |\varphi_1^{\dagger} \varphi_2|^2 |\varphi_1|^2 + c_{(1221)2} |\varphi_1^{\dagger} \varphi_2|^2 |\varphi_2|^2 \nn\\
&&+ c_{(1212)1} ((\varphi_1^{\dagger} \varphi_2)^2 + h.c.) |\varphi_1|^2 
+ c_{(1212)2}  ((\varphi_1^{\dagger} \varphi_2)^2 + h.c.) |\varphi_2|^2 \nn\\
&&+ \textcolor{blue}{c_{(1221)12} |\varphi_1^{\dagger} \varphi_2|^2 (\varphi_1^{\dagger} \varphi_2 + h.c.)} 
+ \textcolor{blue}{c_{11(12)} |\varphi_1|^4 (\varphi_1^{\dagger} \varphi_2 + h.c.)} \nn\\
&&+ \textcolor{blue}{c_{22(12)} |\varphi_2|^4 (\varphi_1^{\dagger} \varphi_2 + h.c.)} 
+ \textcolor{blue}{c_{12(12)} |\varphi_1|^2 |\varphi_2|^2 (\varphi_1^{\dagger} \varphi_2 + h.c.)}\nn\\ 
&&+ \textcolor{blue}{c_{121212} (\varphi_1^{\dagger} \varphi_2 + h.c.)^3}  \Big].
\end{eqnarray*}
The minimisation conditions of this potential are: 
\begin{eqnarray*}
&&\frac{3}{4}v_1^4 c_{111} + \frac{v_1^2 v_2^2}{2}c_{112} + \frac{v_2^4}{4}c_{122}+ v_1^2 v_2^2 c_{(1212)1} + \frac{v_2^4}{2}c_{(1212)2}+ \frac{v_1^2 v_2^2}{2}c_{(1221)1}  + \frac{v_2^4}{4}c_{(1221)2} \nn\\
&&+ \frac{3}{4}v_1 v_2^3 c_{(1221)12} + \frac{5}{4}v_1^3 v_2 c_{11(12)} + \frac{3}{4}v_1 v_2^3 c_{12(12)} + \frac{v_2^5}{4 v_1}c_{22(12)} + 3 v_1 v_2^3 c_{121212}  = 0 \, ,\\
\text{and}\\
&&\frac{3}{4}v_{2}^4 c_{222} + \frac{v_1^4}{4}c_{112} + \frac{v_1^2 v_2^2}{2}c_{122} +\frac{v_1^4}{2}c_{(1212)1} + v_1^2 v_2^2 c_{(1212)2}+ \frac{v_1^4}{4} c_{(1221)1} + \frac{v_1^2 v_2^2}{2}c_{(1221)2} \nn\\
&&+\frac{3}{4}v_1^3 v_2 c_{(1221)12} + \frac{5}{4} v_1 v_2^3 c_{22(12)} + \frac{3}{4} v_1^3 v_2 c_{(12)12} + 3 v_1^3 v_2 c_{121212} = 0\, .
\end{eqnarray*}
\section{Expressions for scalar mass matrices}
\label{appendix4}
We define the total squared mass matrix $\mathcal{M}^2_{\rho} = m^2_{\rho} + \Delta m^{2}_{\rho}$ for the neutral scalars as: 
\begin{equation*}
\mathcal{L}_{mass} \supset \frac{1}{2}(\rho_1,\rho_2)^T \, \mathcal{M}^2_{\rho} (\rho_1,\rho_2).
\end{equation*}
The mass of heavier scalar and its mixing with the SM-like scalar can be derived according to eqn (3.10), given the following $(1,1)$ and $(1,2)$ elements of the full mass matrix,
\begin{eqnarray*}
\mathcal{M}_{11\rho}^2 &=& -\frac{\Delta_{11\rho}}{2f^2} \Big(m_{12}^2\frac{v_2}{v_1} + \l_1 v_1^2 + \frac{3}{2}\l_6 v_1 v_2 - \frac{1}{2}\l_7 \frac{v_2^3}{v_1} \Big) \nn\\
  &&- \frac{\Delta_{12\rho}}{4f^2} \Big[-m_{12}^2 + (\l_3+\l_4+\l_5)v_1 v_2 + \frac{3}{2}  (\l_6 v_1^2 +  \l_7 v_2^2) \Big], \nn\\
 \mathcal{M}_{12\rho}^2 &=& -\frac{\Delta_{12\rho}}{8f^2} \Big[ m_{12}^2 \frac{v^2}{v_1 v_2} + \l_1 v_1^2 + \l_2 v_2^2 + \frac{3}{2}(\l_6 + \l_7)v_1 v_2 - \frac{1}{2}\Big(\l_6\frac{v_1^3}{v_2} + \l_7 \frac{v_2^3}{v_1}\Big) \Big] \nn\\
  &&- \frac{\Delta_{11\rho} + \Delta_{22\rho}}{4f^2} \Big[-m_{12}^2 + (\l_3+\l_4+\l_5)v_1 v_2 + \frac{3}{2}  (\l_6 v_1^2 +  \l_7 v_2^2) \Big].
\end{eqnarray*}

Mass of the physical psuedoscalar,
\begin{eqnarray*}
M_{A}^2 &=& (M_{A}^2)^{tree}\Big[1 + \frac{1}{4f^2}\Big(2 s_{\b} c_{\b} \Delta_{12\eta} - 2 s_{\b}^2 \Delta_{11\eta} - 2 c_{\b}^2 \Delta_{22\eta} \Big) \Big] \nn\\
&&- \frac{v^4}{f^2} \Big[c_{\b}^2 c_{(1212)1} + s_{\b}^2 c_{(1212)2} + \frac{1}{4} \frac{1}{\tan \b} c_{\b}^2 c_{11(12)} + \frac{1}{4} \tan \beta s_{\b}^2 c_{22(12)} \nn\\
&&+ s_{\b} c_{\b} \Big(\frac{1}{4}c_{12(12)} + \frac{1}{4}c_{(1221)12}  + 3 c_{121212} \Big) \Big].
\end{eqnarray*}

Mass of the physical charged scalar, 
\begin{eqnarray*}
M_{H^{\pm}}^2 &=& (M_{H^{\pm}}^2)^{tree} - \frac{v^4}{f^2}\Big[\frac{1}{2}\Big(c_{\b}^2 (c_{(1212)1} + \frac{1}{2}c_{(1221)1}) + s_{\b}^2 (c_{(1212)2} + \frac{1}{2}c_{(1221)2})\Big)\\ 
&&+ s_{\b} c_{\b} \Big(\frac{3}{4}c_{(1221)12} + \frac{1}{4}c_{12(12)} + 3 c_{121212} \Big)  + \frac{1}{4} \frac{1}{\tan \beta} c_{\b}^2c_{11(12)} + \frac{1}{4} \tan \beta s_{\b}^2 c_{22(12)}  \Big].
\end{eqnarray*}
%\label{appendix1}
 
\section{Anomalous TGVs and \textbf{\textit{hVV}}}
\label{appendix5}
The definition of anomalous TGVs are given as: 
\bea
\mathcal{L}_{TGV} &=& i g \cos \theta_{w} \delta g_{Z}^{1} Z^{\m}(W^{-\n}W^{+}_{\m\n} - W^{+ \n}W^{-}_{\m\n}) +  i g (\delta \kappa_{Z} \cos \theta_{w} Z^{\m\n} + \delta \kappa_{\g} \sin \theta_{w} F^{\m\n})W^{-}_{\m}W^{+}_{\n} \nn\\
&&+ \frac{i g}{m_{W}^2} (\l_{Z}\cos \theta_{w} Z^{\m\n} + \lambda_{\gamma} \sin \theta_{w} F^{\m\n}) W^{-\rho}_{\n}W^{+}_{\rho\m}. \nn 
\eea  
Operators of type $\varphi^2 D^2 X$ lead to these kinds of effective couplings of Higgs with vector bosons:
\bea
\mathcal{L}_{hVV} &\supset & \Big(c_{WW} W_{\m\n}^{+} W^{-\m\n} + \frac{c_{ZZ}}{2}Z_{\m\n}Z^{\m\n} + c_{Z\gamma} F_{\m\n}Z^{\m\n} + \frac{c_{\gamma\gamma}}{2}F_{\m\n}F^{\m\n} \Big)\frac{h}{v}\nn\\
&&+ \Big(c_{W\partial W} (W_{\n}^{-}D_{\m}W^{+\m\n} + h.c.) + c_{Z\partial Z} Z_{\m} \partial_{\n} Z^{\m\n} \Big)\frac{h}{v}.\nn
\eea


\begin{thebibliography}{99}
  
 
%\cite{Trodden:1998ym}
\bibitem{Trodden:1998ym}
  M.~Trodden,
  %``Electroweak baryogenesis,''
  Rev.\ Mod.\ Phys.\  {\bf 71} (1999) 1463
  doi:10.1103/RevModPhys.71.1463
  [hep-ph/9803479].
  %%CITATION = doi:10.1103/RevModPhys.71.1463;%%
  %315 citations counted in INSPIRE as of 07 Jan 2017


%\cite{Crivellin:2012ye}
\bibitem{Crivellin:2012ye}
  A.~Crivellin, C.~Greub and A.~Kokulu,
  %``Explaining $B\to D\tau\nu$, $B\to D^*\tau\nu$ and $B\to \tau\nu$ in a 2HDM of type III,''
  Phys.\ Rev.\ D {\bf 86} (2012) 054014
  doi:10.1103/PhysRevD.86.054014
  [arXiv:1206.2634 [hep-ph]].
  %%CITATION = doi:10.1103/PhysRevD.86.054014;%%
  %149 citations counted in INSPIRE as of 07 Jan 2017
  
  %\cite{Celis:2012dk}
\bibitem{Celis:2012dk}
  A.~Celis, M.~Jung, X.~Q.~Li and A.~Pich,
  %``Sensitivity to charged scalars in $\boldsymbol{B\to D^{(*)}\tau\nu_\tau}$ and $\boldsymbol{B\to\tau\nu_\tau}$ decays,''
  JHEP {\bf 1301} (2013) 054
  doi:10.1007/JHEP01(2013)054
  [arXiv:1210.8443 [hep-ph]].
  %%CITATION = doi:10.1007/JHEP01(2013)054;%%
  %124 citations counted in INSPIRE as of 07 Jan 2017
  
  %\cite{Crivellin:2015hha}
\bibitem{Crivellin:2015hha}
  A.~Crivellin, J.~Heeck and P.~Stoffer,
  %``A perturbed lepton-specific two-Higgs-doublet model facing experimental hints for physics beyond the Standard Model,''
  Phys.\ Rev.\ Lett.\  {\bf 116} (2016) no.8,  081801
  doi:10.1103/PhysRevLett.116.081801
  [arXiv:1507.07567 [hep-ph]].
  %%CITATION = doi:10.1103/PhysRevLett.116.081801;%%
  %61 citations counted in INSPIRE as of 07 Jan 2017
  
  
  %\cite{Giudice:2007fh}
\bibitem{Giudice:2007fh}
  G.~F.~Giudice, C.~Grojean, A.~Pomarol and R.~Rattazzi,
  %``The Strongly-Interacting Light Higgs,''
  JHEP {\bf 0706} (2007) 045
  doi:10.1088/1126-6708/2007/06/045
  [hep-ph/0703164].
  %%CITATION = doi:10.1088/1126-6708/2007/06/045;%%
  %584 citations counted in INSPIRE as of 07 Jan 2017  
  
  
%\cite{Buchmuller:1985jz}
\bibitem{Buchmuller:1985jz} 
  W.~Buchmuller and D.~Wyler,
  %``Effective Lagrangian Analysis of New Interactions and Flavor Conservation,''
  Nucl.\ Phys.\ B {\bf 268}, 621 (1986)
  doi:10.1016/0550-3213(86)90262-2.
  %%CITATION = doi:10.1016/0550-3213(86)90262-2;%%
  %1209 citations counted in INSPIRE as of 13 Jan 2017  
  
%\cite{Leung:1984ni}
\bibitem{Leung:1984ni}
  C.~N.~Leung, S.~T.~Love and S.~Rao,
  %``Low-Energy Manifestations of a New Interaction Scale: Operator Analysis,''
  Z.\ Phys.\ C {\bf 31} (1986) 433.
  doi:10.1007/BF01588041.
  %%CITATION = doi:10.1007/BF01588041;%%
  %266 citations counted in INSPIRE as of 07 May 2017    
  
  %\cite{Hagiwara:1993ck}
\bibitem{Hagiwara:1993ck}
  K.~Hagiwara, S.~Ishihara, R.~Szalapski and D.~Zeppenfeld,
  %``Low-energy effects of new interactions in the electroweak boson sector,''
  Phys.\ Rev.\ D {\bf 48} (1993) 2182.
  doi:10.1103/PhysRevD.48.2182.
  %%CITATION = doi:10.1103/PhysRevD.48.2182;%%
  %548 citations counted in INSPIRE as of 18 Jun 2017

  
%\cite{Grzadkowski:2010es}
\bibitem{Grzadkowski:2010es}
  B.~Grzadkowski, M.~Iskrzynski, M.~Misiak and J.~Rosiek,
  %``Dimension-Six Terms in the Standard Model Lagrangian,''
  JHEP {\bf 1010} (2010) 085
  doi:10.1007/JHEP10(2010)085
  [arXiv:1008.4884 [hep-ph]].
  %%CITATION = doi:10.1007/JHEP10(2010)085;%%
  %500 citations counted in INSPIRE as of 11 Jan 2017  
  

  
   


  

%\cite{Agashe:2004rs}
\bibitem{Agashe:2004rs} 
  K.~Agashe, R.~Contino and A.~Pomarol,
  %``The Minimal composite Higgs model,''
  Nucl.\ Phys.\ B {\bf 719}, 165 (2005)
  doi:10.1016/j.nuclphysb.2005.04.035
  [hep-ph/0412089].
  %%CITATION = doi:10.1016/j.nuclphysb.2005.04.035;%%
  %909 citations counted in INSPIRE as of 12 Jan 2017   
   
   %\cite{ArkaniHamed:2002qy}
\bibitem{ArkaniHamed:2002qy}
  N.~Arkani-Hamed, A.~G.~Cohen, E.~Katz and A.~E.~Nelson,
  %``The Littlest Higgs,''
  JHEP {\bf 0207} (2002) 034
  doi:10.1088/1126-6708/2002/07/034
  [hep-ph/0206021].
  %%CITATION = doi:10.1088/1126-6708/2002/07/034;%%
  %1041 citations counted in INSPIRE as of 10 May 2017
  
%\cite{Falkowski:2015fla}
\bibitem{Falkowski:2015fla}
  A.~Falkowski,
  %``Effective field theory approach to LHC Higgs data,''
  Pramana {\bf 87} (2016) no.3,  39
  doi:10.1007/s12043-016-1251-5
  [arXiv:1505.00046 [hep-ph]].
  %%CITATION = doi:10.1007/s12043-016-1251-5;%%
  %40 citations counted in INSPIRE as of 29 Jun 2017  
  
  %\cite{Brivio:2017vri}
\bibitem{Brivio:2017vri}
  I.~Brivio and M.~Trott,
  %``The Standard Model as an Effective Field Theory,''
  arXiv:1706.08945 [hep-ph].
  %%CITATION = ARXIV:1706.08945;%%
  
%\cite{Mrazek:2011iu}
\bibitem{Mrazek:2011iu} 
  J.~Mrazek, A.~Pomarol, R.~Rattazzi, M.~Redi, J.~Serra and A.~Wulzer,
  %``The Other Natural Two Higgs Doublet Model,''
  Nucl.\ Phys.\ B {\bf 853}, 1 (2011)
  doi:10.1016/j.nuclphysb.2011.07.008
  [arXiv:1105.5403 [hep-ph]].
  %%CITATION = doi:10.1016/j.nuclphysb.2011.07.008;%%
  %122 citations counted in INSPIRE as of 12 Jan 2017  
  
%\cite{DeCurtis:2016scv}
\bibitem{DeCurtis:2016scv}
  S.~De Curtis, S.~Moretti, K.~Yagyu and E.~Yildirim,
  %``Perturbative unitarity bounds in composite two-Higgs doublet models,''
  Phys.\ Rev.\ D {\bf 94} (2016) no.5,  055017
  doi:10.1103/PhysRevD.94.055017
  [arXiv:1602.06437 [hep-ph]].
  %%CITATION = doi:10.1103/PhysRevD.94.055017;%%
  %6 citations counted in INSPIRE as of 08 May 2017
  
  %\cite{DeCurtis:2016gly}
\bibitem{DeCurtis:2016gly}
  S.~De Curtis, S.~Moretti, K.~Yagyu and E.~Yildirim,
  %``Theory and Phenomenology of Composite 2-Higgs Doublet Models,''
  arXiv:1612.05125 [hep-ph].
  %%CITATION = ARXIV:1612.05125;%%
  %1 citations counted in INSPIRE as of 08 May 2017
  
  %\cite{DeCurtis:2017gzi}
\bibitem{DeCurtis:2017gzi}
  S.~De Curtis, S.~Moretti, K.~Yagyu and E.~Yildirim,
  %``Single and Double SM-like Higgs Boson Production at Future Electron-Positron Colliders in C2HDMs,''
  arXiv:1702.07260 [hep-ph].
  %%CITATION = ARXIV:1702.07260;%%  
  
  %\cite{Brown:2010ke}
\bibitem{Brown:2010ke}
  T.~Brown, C.~Frugiuele and T.~Gregoire,
  %``UV friendly T-parity in the SU(6)/Sp(6) little Higgs model,''
  JHEP {\bf 1106} (2011) 108
  doi:10.1007/JHEP06(2011)108
  [arXiv:1012.2060 [hep-ph]].
  %%CITATION = doi:10.1007/JHEP06(2011)108;%%
  %11 citations counted in INSPIRE as of 30 Jun 2017
  
  %\cite{Gopalakrishna:2015dkt}
\bibitem{Gopalakrishna:2015dkt}
  S.~Gopalakrishna, T.~S.~Mukherjee and S.~Sadhukhan,
  %``Status and Prospects of the Two-Higgs-Doublet SU(6)/Sp(6) little-Higgs Model and the Alignment Limit,''
  Phys.\ Rev.\ D {\bf 94} (2016) no.1,  015034
  doi:10.1103/PhysRevD.94.015034
  [arXiv:1512.05731 [hep-ph]].
  %%CITATION = doi:10.1103/PhysRevD.94.015034;%%
  %6 citations counted in INSPIRE as of 08 May 2017 
  
  %\cite{Fonseca:2015gva}
\bibitem{Fonseca:2015gva}
  N.~Fonseca, R.~Zukanovich Funchal, A.~Lessa and L.~Lopez-Honorez,
  %``Dark Matter Constraints on Composite Higgs Models,''
  JHEP {\bf 1506} (2015) 154
  doi:10.1007/JHEP06(2015)154
  [arXiv:1501.05957 [hep-ph]].
  %%CITATION = doi:10.1007/JHEP06(2015)154;%%
  %12 citations counted in INSPIRE as of 08 May 2017
  
  
%\cite{Carmona:2015haa}
\bibitem{Carmona:2015haa}
  A.~Carmona and M.~Chala,
  %``Composite Dark Sectors,''
  JHEP {\bf 1506} (2015) 105
  doi:10.1007/JHEP06(2015)105
  [arXiv:1504.00332 [hep-ph]].
  %%CITATION = doi:10.1007/JHEP06(2015)105;%%
  %27 citations counted in INSPIRE as of 08 May 2017
  
  
  %\cite{Chala:2017sjk}
\bibitem{Chala:2017sjk}
  M.~Chala, G.~Durieux, C.~Grojean, L.~de Lima and O.~Matsedonskyi,
  %``Minimally extended SILH,''
  arXiv:1703.10624 [hep-ph].
  %%CITATION = ARXIV:1703.10624;%%
  %1 citations counted in INSPIRE as of 08 May 2017


%\cite{DiazCruz:2001tn}
\bibitem{DiazCruz:2001tn}
  J.~L.~Diaz-Cruz, J.~Hernandez-Sanchez and J.~J.~Toscano,
  %``An Effective Lagrangian description of charged Higgs decays $H^{+} \to W^{+} \gamma$, $W^{+} Z$ and $W^{+}$ h0,''
  Phys.\ Lett.\ B {\bf 512} (2001) 339
  doi:10.1016/S0370-2693(01)00703-1
  [hep-ph/0106001].
  %%CITATION = doi:10.1016/S0370-2693(01)00703-1;%%
  %33 citations counted in INSPIRE as of 29 Jun 2017  
  
  %\cite{Kikuta:2015pya}
\bibitem{Kikuta:2015pya}
  Y.~Kikuta and Y.~Yamamoto,
  %``Derivative interactions and perturbative UV contributions in N Higgs Doublet Models,''
  Eur.\ Phys.\ J.\ C {\bf 76} (2016) no.5,  297
  doi:10.1140/epjc/s10052-016-4128-3
  [arXiv:1510.05540 [hep-ph]].
  %%CITATION = doi:10.1140/epjc/s10052-016-4128-3;%%
  %1 citations counted in INSPIRE as of 08 May 2017
  
  %\cite{Kikuta:2011ew}
\bibitem{Kikuta:2011ew}
  Y.~Kikuta, Y.~Okada and Y.~Yamamoto,
  %``Structure of dimension-six derivative interactions in pseudo Nambu-Goldstone N Higgs doublet models,''
  Phys.\ Rev.\ D {\bf 85} (2012) 075021
  doi:10.1103/PhysRevD.85.075021
  [arXiv:1111.2120 [hep-ph]].
  %%CITATION = doi:10.1103/PhysRevD.85.075021;%%
  %3 citations counted in INSPIRE as of 08 May 2017
  
  
%\cite{Kikuta:2012tf}
\bibitem{Kikuta:2012tf}
  Y.~Kikuta and Y.~Yamamoto,
  %``Perturbative unitarity of Higgs derivative interactions,''
  PTEP {\bf 2013} (2013) 053B05
  doi:10.1093/ptep/ptt030
  [arXiv:1210.5674 [hep-ph]].
  %%CITATION = doi:10.1093/ptep/ptt030;%%
  %9 citations counted in INSPIRE as of 08 May 2017
  
  %\cite{Crivellin:2016ihg}
\bibitem{Crivellin:2016ihg}
  A.~Crivellin, M.~Ghezzi and M.~Procura,
  %``Effective Field Theory with Two Higgs Doublets,''
  JHEP {\bf 1609} (2016) 160
  doi:10.1007/JHEP09(2016)160
  [arXiv:1608.00975 [hep-ph]].
  %%CITATION = doi:10.1007/JHEP09(2016)160;%%
  %2 citations counted in INSPIRE as of 07 Jan 2017
   

  
%\cite{Ginzburg:2004vp}
\bibitem{Ginzburg:2004vp}
  I.~F.~Ginzburg and M.~Krawczyk,
  %``Symmetries of two Higgs doublet model and CP violation,''
  Phys.\ Rev.\ D {\bf 72} (2005) 115013
  doi:10.1103/PhysRevD.72.115013
  [hep-ph/0408011].
  %%CITATION = doi:10.1103/PhysRevD.72.115013;%%
  %148 citations counted in INSPIRE as of 24 Jan 2017  
  
%\cite{Ginzburg:2008kr}
\bibitem{Ginzburg:2008kr}
  I.~F.~Ginzburg,
  %``Necessity of mixed kinetic term in the description of general system with identical scalar fields,''
  Phys.\ Lett.\ B {\bf 682} (2009) 61
  doi:10.1016/j.physletb.2009.10.071
  [arXiv:0810.1546 [hep-ph]].
  %%CITATION = doi:10.1016/j.physletb.2009.10.071;%%
  %5 citations counted in INSPIRE as of 14 Feb 2017  
  
%\cite{Politzer:1980me}
\bibitem{Politzer:1980me}
  H.~D.~Politzer,
  %``Power Corrections at Short Distances,''
  Nucl.\ Phys.\ B {\bf 172} (1980) 349
  doi:10.1016/0550-3213(80)90172-8.
  %%CITATION = doi:10.1016/0550-3213(80)90172-8;%%
  %300 citations counted in INSPIRE as of 10 Feb 2017  
  
  
%\cite{Barbieri:2004qk}
\bibitem{Barbieri:2004qk}
  R.~Barbieri, A.~Pomarol, R.~Rattazzi and A.~Strumia,
  %``Electroweak symmetry breaking after LEP-1 and LEP-2,''
  Nucl.\ Phys.\ B {\bf 703} (2004) 127
  doi:10.1016/j.nuclphysb.2004.10.014
  [hep-ph/0405040].
  %%CITATION = doi:10.1016/j.nuclphysb.2004.10.014;%%
  %481 citations counted in INSPIRE as of 19 Jun 2017
  
  %\cite{Elias-Miro:2013mua}
\bibitem{Elias-Miro:2013mua} 
  J.~Elias-Miro, J.~R.~Espinosa, E.~Masso and A.~Pomarol,
  %``Higgs windows to new physics through d=6 operators: constraints and one-loop anomalous dimensions,''
  JHEP {\bf 1311}, 066 (2013)
  doi:10.1007/JHEP11(2013)066
  [arXiv:1308.1879 [hep-ph]].
  %%CITATION = doi:10.1007/JHEP11(2013)066;%%
  %132 citations counted in INSPIRE as of 11 Jan 2017   
  
    %\cite{Carena:2013ooa}
\bibitem{Carena:2013ooa}
  M.~Carena, I.~Low, N.~R.~Shah and C.~E.~M.~Wagner,
  %``Impersonating the Standard Model Higgs Boson: Alignment without Decoupling,''
  JHEP {\bf 1404} (2014) 015
  doi:10.1007/JHEP04(2014)015
  [arXiv:1310.2248 [hep-ph]].
  %%CITATION = doi:10.1007/JHEP04(2014)015;%%
  %86 citations counted in INSPIRE as of 16 Apr 2017  
  
%\cite{Hagiwara:1986vm}
\bibitem{Hagiwara:1986vm}
  K.~Hagiwara, R.~D.~Peccei, D.~Zeppenfeld and K.~Hikasa,
  %``Probing the Weak Boson Sector in e+ e- ---> W+ W-,''
  Nucl.\ Phys.\ B {\bf 282} (1987) 253
  doi:10.1016/0550-3213(87)90685-7.
  %%CITATION = doi:10.1016/0550-3213(87)90685-7;%%
  %1239 citations counted in INSPIRE as of 29 Jun 2017  
  
  %\cite{LEPEWWG}
  \bibitem{LEPEWWG}
The LEP collaborations ALEPH, DELPHI, L3, OPAL, and the LEP TGC Working
Group, LEPEWWG/TGC/2003-01.  
  
  
%\cite{Falkowski:2015jaa}
\bibitem{Falkowski:2015jaa}
  A.~Falkowski, M.~Gonzalez-Alonso, A.~Greljo and D.~Marzocca,
  %``Global constraints on anomalous triple gauge couplings in effective field theory approach,''
  Phys.\ Rev.\ Lett.\  {\bf 116} (2016) no.1,  011801
  doi:10.1103/PhysRevLett.116.011801
  [arXiv:1508.00581 [hep-ph]].
  %%CITATION = doi:10.1103/PhysRevLett.116.011801;%%
  %37 citations counted in INSPIRE as of 28 Jun 2017  
  
%\cite{Maksymyk:1993zm}
\bibitem{Maksymyk:1993zm}
  I.~Maksymyk, C.~P.~Burgess and D.~London,
  %``Beyond S, T and U,''
  Phys.\ Rev.\ D {\bf 50} (1994) 529
  doi:10.1103/PhysRevD.50.529
  [hep-ph/9306267].
  %%CITATION = doi:10.1103/PhysRevD.50.529;%%
  %161 citations counted in INSPIRE as of 21 Jun 2017    
    
   
  %\cite{Peskin:1991sw}
\bibitem{Peskin:1991sw}
  M.~E.~Peskin and T.~Takeuchi,
  %``Estimation of oblique electroweak corrections,''
  Phys.\ Rev.\ D {\bf 46} (1992) 381
  doi:10.1103/PhysRevD.46.381.
  %%CITATION = doi:10.1103/PhysRevD.46.381;%%
  %1821 citations counted in INSPIRE as of 16 Apr 2017
  

  
  %\cite{Baak:2014ora}
\bibitem{Baak:2014ora}
  M.~Baak {\it et al.} [Gfitter Group],
  %``The global electroweak fit at NNLO and prospects for the LHC and ILC,''
  Eur.\ Phys.\ J.\ C {\bf 74} (2014) 3046
  doi:10.1140/epjc/s10052-014-3046-5
  [arXiv:1407.3792 [hep-ph]].
  %%CITATION = doi:10.1140/epjc/s10052-014-3046-5;%%
  %297 citations counted in INSPIRE as of 08 May 2017 


%\cite{Burgess:1993vc}
\bibitem{Burgess:1993vc}
  C.~P.~Burgess, S.~Godfrey, H.~Konig, D.~London and I.~Maksymyk,
  %``Model independent global constraints on new physics,''
  Phys.\ Rev.\ D {\bf 49} (1994) 6115
  doi:10.1103/PhysRevD.49.6115
  [hep-ph/9312291].
  %%CITATION = doi:10.1103/PhysRevD.49.6115;%%
  %179 citations counted in INSPIRE as of 25 May 2017 


%\cite{Haber:2010bw}
\bibitem{Haber:2010bw}
  H.~E.~Haber and D.~O'Neil,
  %``Basis-independent methods for the two-Higgs-doublet model III: The CP-conserving limit, custodial symmetry, and the oblique parameters S, T, U,''
  Phys.\ Rev.\ D {\bf 83} (2011) 055017
  doi:10.1103/PhysRevD.83.055017
  [arXiv:1011.6188 [hep-ph]].
  %%CITATION = doi:10.1103/PhysRevD.83.055017;%%
  %80 citations counted in INSPIRE as of 22 Jun 2017


%\cite{Kennedy:1988sn}
\bibitem{Kennedy:1988sn}
  D.~C.~Kennedy and B.~W.~Lynn,
  %``Electroweak Radiative Corrections with an Effective Lagrangian: Four Fermion Processes,''
  Nucl.\ Phys.\ B {\bf 322} (1989) 1.
  doi:10.1016/0550-3213(89)90483-5.
  %%CITATION = doi:10.1016/0550-3213(89)90483-5;%%
  %428 citations counted in INSPIRE as of 21 Jun 2017

 

%\cite{Burgess:1994zp}
\bibitem{Burgess:1994zp}
  C.~P.~Burgess,
  %``The Effective use of precision electroweak measurements,''
  Pramana {\bf 45} (1995) S47
  doi:10.1007/BF02907965
  [hep-ph/9411257].
  %%CITATION = doi:10.1007/BF02907965;%%
  %10 citations counted in INSPIRE as of 22 Jun 2017 

%\cite{Aoki:2009ha}
\bibitem{Aoki:2009ha}
  M.~Aoki, S.~Kanemura, K.~Tsumura and K.~Yagyu,
  %``Models of Yukawa interaction in the two Higgs doublet model, and their collider phenomenology,''
  Phys.\ Rev.\ D {\bf 80} (2009) 015017
  doi:10.1103/PhysRevD.80.015017
  [arXiv:0902.4665 [hep-ph]].
  %%CITATION = doi:10.1103/PhysRevD.80.015017;%%
  %206 citations counted in INSPIRE as of 10 May 2017  
  
  
%\cite{Eberhardt:2013uba}
\bibitem{Eberhardt:2013uba}
  O.~Eberhardt, U.~Nierste and M.~Wiebusch,
  %``Status of the two-Higgs-doublet model of type II,''
  JHEP {\bf 1307} (2013) 118
  doi:10.1007/JHEP07(2013)118
  [arXiv:1305.1649 [hep-ph]].
  %%CITATION = doi:10.1007/JHEP07(2013)118;%%
  %106 citations counted in INSPIRE as of 16 Apr 2017  
  
  %\cite{Haber:2015pua}
\bibitem{Haber:2015pua}
  H.~E.~Haber and O.~St{\"a}l,
  %``New LHC benchmarks for the $\mathcal{CP}$ -conserving two-Higgs-doublet model,''
  Eur.\ Phys.\ J.\ C {\bf 75} (2015) no.10,  491
   Erratum: [Eur.\ Phys.\ J.\ C {\bf 76} (2016) no.6,  312]
  doi:10.1140/epjc/s10052-015-3697-x, 10.1140/epjc/s10052-016-4151-4
  [arXiv:1507.04281 [hep-ph]].
  %%CITATION = doi:10.1140/epjc/s10052-015-3697-x, 10.1140/epjc/s10052-016-4151-4;%%
  %36 citations counted in INSPIRE as of 16 Apr 2017
 
 %\cite{Belusca-Maito:2016dqe}
\bibitem{Belusca-Maito:2016dqe}
  H.~Bélusca-Maïto, A.~Falkowski, D.~Fontes, J.~C.~Romão and J.~P.~Silva,
  %``Higgs EFT for 2HDM and beyond,''
  Eur.\ Phys.\ J.\ C {\bf 77} (2017) no.3,  176
  doi:10.1140/epjc/s10052-017-4745-5
  [arXiv:1611.01112 [hep-ph]].
  %%CITATION = doi:10.1140/epjc/s10052-017-4745-5;%%
  %2 citations counted in INSPIRE as of 16 Apr 2017 
  
%\cite{Appelquist:1974tg}
\bibitem{Appelquist:1974tg}
  T.~Appelquist and J.~Carazzone,
  %``Infrared Singularities and Massive Fields,''
  Phys.\ Rev.\ D {\bf 11} (1975) 2856.
  doi:10.1103/PhysRevD.11.2856.
  %%CITATION = doi:10.1103/PhysRevD.11.2856;%%
  %1583 citations counted in INSPIRE as of 21 Jun 2017  
  
  
  %\cite{Gunion:2002zf}
\bibitem{Gunion:2002zf}
  J.~F.~Gunion and H.~E.~Haber,
  %``The CP conserving two Higgs doublet model: The Approach to the decoupling limit,''
  Phys.\ Rev.\ D {\bf 67} (2003) 075019
  doi:10.1103/PhysRevD.67.075019
  [hep-ph/0207010].
  %%CITATION = doi:10.1103/PhysRevD.67.075019;%%
  %468 citations counted in INSPIRE as of 16 Apr 2017
  
  %\cite{Dev:2014yca}
\bibitem{Dev:2014yca}
  P.~S.~Bhupal Dev and A.~Pilaftsis,
  %``Maximally Symmetric Two Higgs Doublet Model with Natural Standard Model Alignment,''
  JHEP {\bf 1412} (2014) 024
   Erratum: [JHEP {\bf 1511} (2015) 147]
  doi:10.1007/JHEP11(2015)147, 10.1007/JHEP12(2014)024
  [arXiv:1408.3405 [hep-ph]].
  %%CITATION = doi:10.1007/JHEP11(2015)147, 10.1007/JHEP12(2014)024;%%
  %94 citations counted in INSPIRE as of 05 Sep 2017


%\cite{Contino:2013kra}
\bibitem{Contino:2013kra}
  R.~Contino, M.~Ghezzi, C.~Grojean, M.~Muhlleitner and M.~Spira,
  %``Effective Lagrangian for a light Higgs-like scalar,''
  JHEP {\bf 1307} (2013) 035
  doi:10.1007/JHEP07(2013)035
  [arXiv:1303.3876 [hep-ph]].
  %%CITATION = doi:10.1007/JHEP07(2013)035;%%
  %235 citations counted in INSPIRE as of 28 Jun 2017 

%\cite{Djouadi:2005gj}
\bibitem{Djouadi:2005gj}
  A.~Djouadi,
  %``The Anatomy of electro-weak symmetry breaking. II. The Higgs bosons in the minimal supersymmetric model,''
  Phys.\ Rept.\  {\bf 459} (2008) 1
  doi:10.1016/j.physrep.2007.10.005
  [hep-ph/0503173].
  %%CITATION = doi:10.1016/j.physrep.2007.10.005;%%
  %1086 citations counted in INSPIRE as of 25 May 2017 

  
%\cite{Falkowski:2013dza}
\bibitem{Falkowski:2013dza}
  A.~Falkowski, F.~Riva and A.~Urbano,
  %``Higgs at last,''
  JHEP {\bf 1311} (2013) 111
  doi:10.1007/JHEP11(2013)111
  [arXiv:1303.1812 [hep-ph]].
  %%CITATION = doi:10.1007/JHEP11(2013)111;%%
  %205 citations counted in INSPIRE as of 26 May 2017  
  
  
 
  

%\cite{Cahn:1988ru}
\bibitem{Cahn:1988ru}
  R.~N.~Cahn,
  %``The Higgs Boson,''
  Rept.\ Prog.\ Phys.\  {\bf 52} (1989) 389.
  doi:10.1088/0034-4885/52/4/001.
  %%CITATION = doi:10.1088/0034-4885/52/4/001;%%
  %77 citations counted in INSPIRE as of 21 May 2017
  
  
%\cite{Altenkamp:2017ldc}
\bibitem{Altenkamp:2017ldc}
  L.~Altenkamp, S.~Dittmaier and H.~Rzehak,
  %``Renormalization schemes for the Two-Higgs-Doublet Model and applications to h -> WW/ZZ -> 4fermions,''
  arXiv:1704.02645 [hep-ph].
  %%CITATION = ARXIV:1704.02645;%%
  %2 citations counted in INSPIRE as of 21 May 2017  
  
  
  
 


%\cite{Dawson:2013bba}
\bibitem{Dawson:2013bba}
  S.~Dawson {\it et al.},
  %``Working Group Report: Higgs Boson,''
  arXiv:1310.8361 [hep-ex].
  %%CITATION = ARXIV:1310.8361;%%
  %288 citations counted in INSPIRE as of 26 May 2017
 
  

   
%\cite{Luty:1997fk}
\bibitem{Luty:1997fk}
  M.~A.~Luty,
  %``Naive dimensional analysis and supersymmetry,''
  Phys.\ Rev.\ D {\bf 57} (1998) 1531
  doi:10.1103/PhysRevD.57.1531
  [hep-ph/9706235].
  %%CITATION = doi:10.1103/PhysRevD.57.1531;%%
  %149 citations counted in INSPIRE as of 08 May 2017  
  
%\cite{Pomarol:2014dya}
\bibitem{Pomarol:2014dya}
  A.~Pomarol,
  %``Higgs Physics,''
  arXiv:1412.4410 [hep-ph].
  %%CITATION = ARXIV:1412.4410;%%
  %25 citations counted in INSPIRE as of 08 May 2017     
   
   
\end{thebibliography}
\end{document}